\documentclass{raa}

\usepackage{graphicx,times}             %for PS/EPS graphics inclusion, new
\input{epsf.sty}                        %for PS/EPS graphics inclusion, old
\input{psfig.sty}                       %for PS/EPS graphics inclusion, old

\begin{document}

   \title{TeV Gamma Ray Astronomy
%\,$^*$
%\footnotetext{$*$ Supported by the National Natural Science Foundation of China.}
}
%   \subtitle{I. Place Your Subtitle Here}

   \volnopage{Vol.0 (200x) No.0, 000--000}      %%preserved for Editor. DOn't remove!
   \setcounter{page}{1}           %%starting page, preserved for Editor. DOn't remove!

   \author{Wei Cui
      \inst{}\mailto{}
%% Please move "\mailto{}" to the corresponding author of the paper.
%% For single author or all the authors from an institute, use "\inst{}" only
%% Here is an example of three authors come from different institutes.
%   \and E. Rodr\'{\i}guez
%      \inst{2}
%   \and M. Breger
%      \inst{3}
      }

   \institute{Department of Physics, Purdue University, 525 Northwestern 
Avenue, West Lafayette, IN 47907, USA\\
             \email{cui@purdue.edu}
%% Please give the E-mail address of the author, to whom future correspondence and
%% offprint requests will be sent. Note to pair \mailto{} with \email{}
%        \and
%             Instituto de Astrofisica de Andalucia, CSIC,
%             P. O. Box 3004, E-18080 Granada, Spain\\
%        \and
%             Astronomisches Institut der Universit\"{a}t Wien, T\"{u}rkenschanzstr. 17,
%             A-1180 Wien, Austria\\
          }

   \date{Received~~2008 month day; accepted~~2008~~month day}

   \abstract{The field of ground-based gamma ray astronomy has enjoyed rapid
growth in recent years. As an increasing number of sources are detected at TeV
energies, the field has matured and become a viable branch of modern 
astronomy. Lying at the uppermost end of the electromagnetic rainbow, TeV 
photons are always preciously few in number but carry essential information 
about the particle acceleration and radiative processes involved in extreme 
astronomical settings. Together with observations at longer wavelengths, 
TeV gamma-ray observations have drastically improved our view of the universe.
In this review, we briefly describe recent progress in the field. We will 
conclude by providing a personal perspective on the future of the field, in 
particular, on the significant roles that China could play to advance this 
young but exciting field. }


   \authorrunning{Wei Cui }            %author_head in even pages
   \titlerunning{TeV Gamma Ray Astronomy}  % title_head in odd pages

   \maketitle
%% The author head (on even pages) and the title head (on odd pages) will be
%% automatically extracted from \author{} and \title{}. Whenever the title is too long,
%% you will be asked to supply a shorter one by inserting either \authorrunning{} or
%% \titlerunning{} before \maketitle. Anyway, you can specify your own heads in advance.
%%
%%
%% Note: In the following text body of your manuscript, please note several differences from
%%       other major journals:
%% (1) \subsection{Please Capitalize the First Letter of Each Notional Word in Subsection Title}
%% (2) Please Capitalize the First Letter of Each Notional Word in all tables' captions

%
%________________________________________________ sections below
%
\section{Introduction}           %% first-level sections will be auto-capitalized
\label{sect:intro}

When one thinks of high energy astronomy, satellites tend to come to mind
right away, because the atmosphere of the Earth is entirely opaque to 
radiation at UV, X-ray, or gamma-ray wavelengths (which is a good thing for 
human vitality). In order to directly detect such radiation, which comes from a 
celestial object, one must therefore place the detector above the atmosphere. 
Although sounding rockets and balloons played a critical role in the early 
days of high energy astronomy, satellites were the ultimate vehicle to launch 
the field into prominence. Over the past several decades, numerous 
breakthroughs in the field have been enabled by space-borne observatories.

However, satellite-based experiments become increasingly ineffective in
detecting gamma rays at increasing energies. When photons reach an energy of
tens of GeV, it is, in fact, problematic to detect them directly, because 
of the practical difficulty in constructing a suitable detector to ``stop'' 
them. This is where ground-based gamma ray experiments come in and contribute. 
At such high energies, there is a unique window of opportunity to do high energy 
astronomy on the ground. Although the gamma rays cannot penetrate the 
atmosphere all the way down to the ground, the consequences of their 
interactions with the atmosphere can be observed and quantified to infer 
their spatial, spectral, and temporal properties. In essence, one uses the 
atmosphere as part of a giant gas detector to register gamma-ray radiation.

\subsection{Experimental Principles}
\label{subsect:principles}

\begin{figure}
   \vspace{2mm}
   \begin{center}
   \hspace{3mm}\psfig{figure=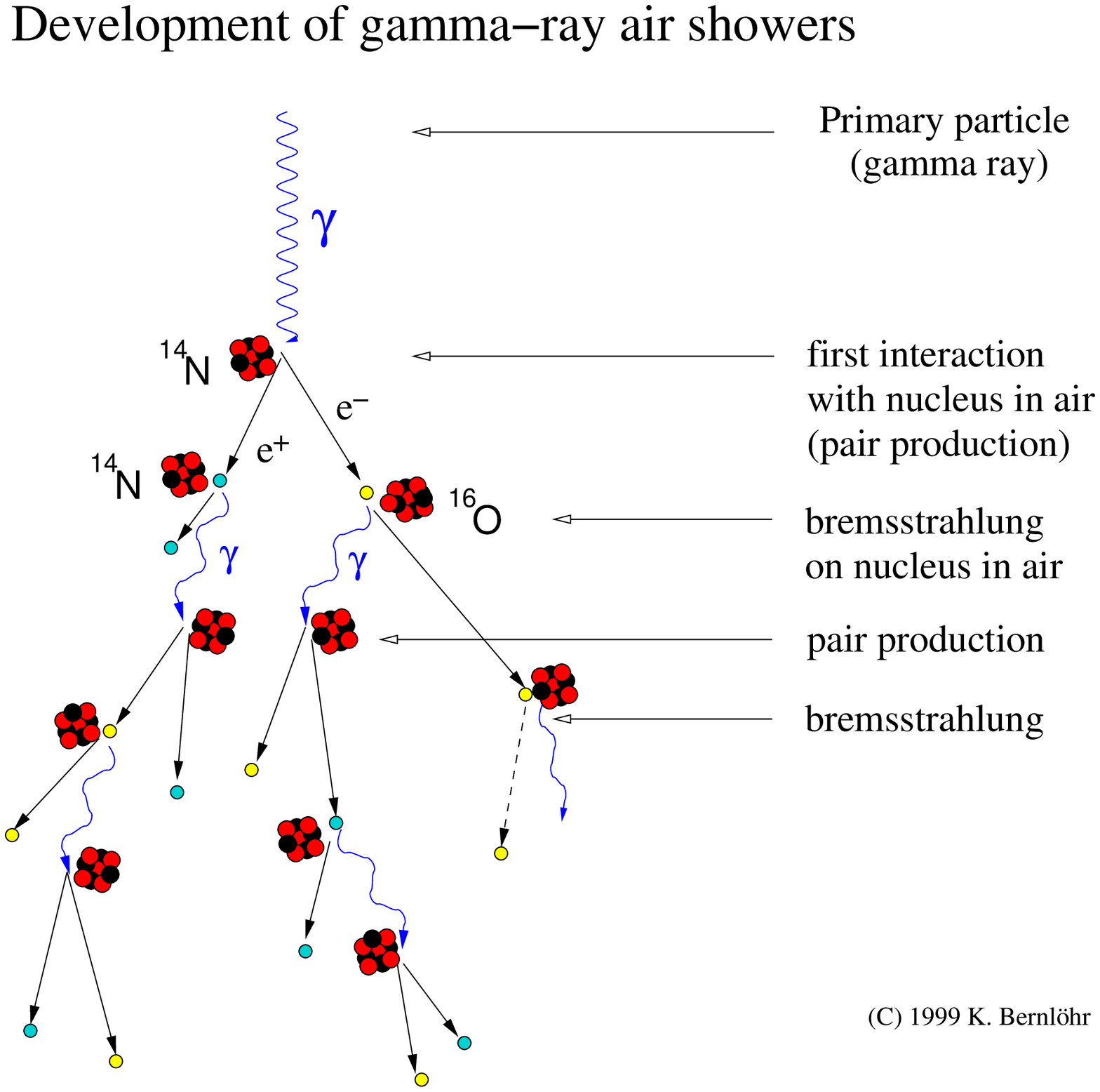,width=70mm,angle=0.0}
   \hspace{3mm}\psfig{figure=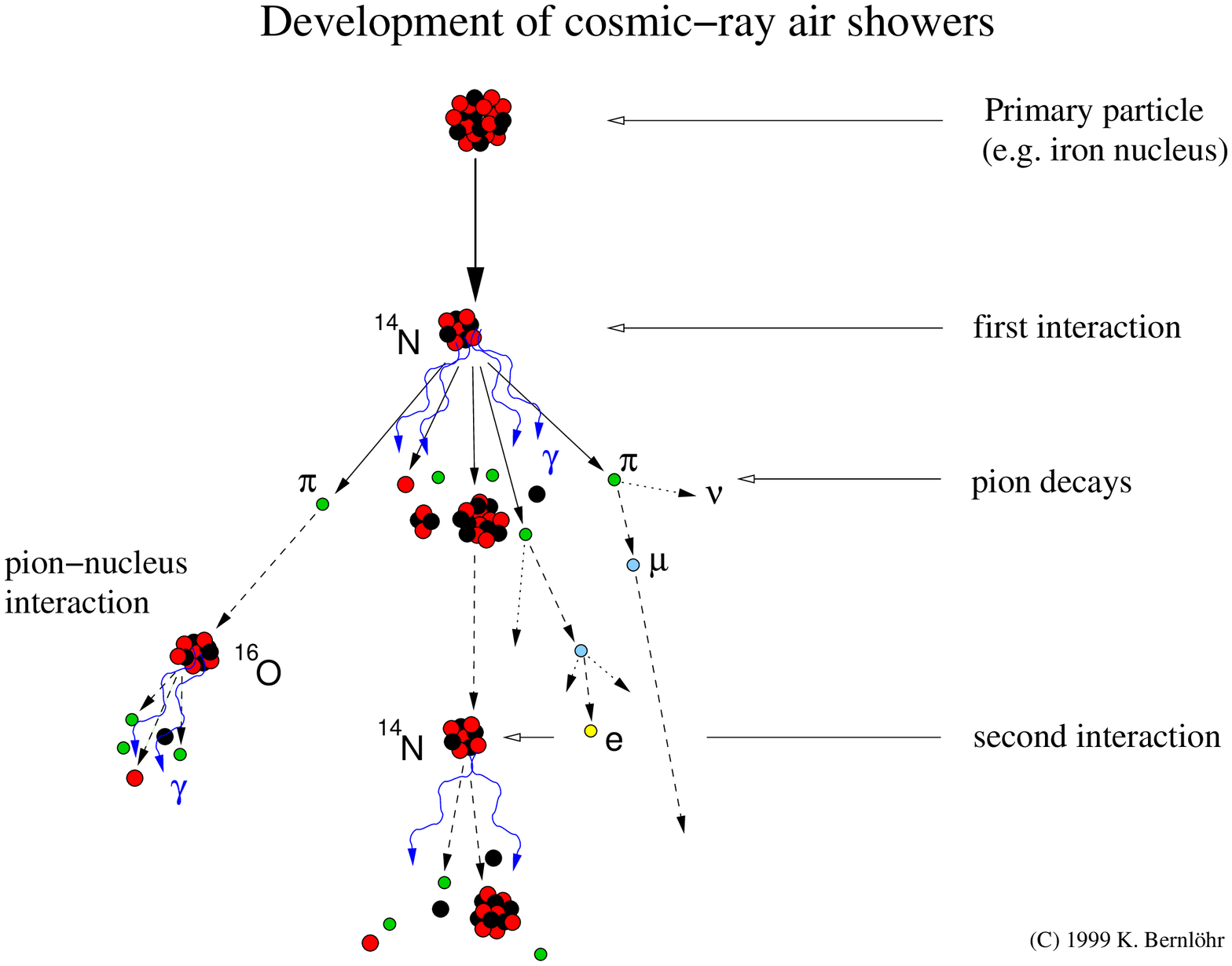,width=70mm,angle=0.0}
%   \parbox{180mm}{{\vspace{2mm} }}
   \caption{Schematic of air shower development. Note the presence of muons
(and neutrinos) associated with hadronic showers. Reproduced with permission
from Konrad Bernl\"{o}hr.  }
   \label{Fig:shower}
   \end{center}
\end{figure}

The interactions between GeV--TeV photons and particles in the atmosphere 
result predominantly in relativistic electron--positron pairs, as illustrated
in Figure~\ref{Fig:shower}. These 
secondary electrons or positrons lose energy mainly by bremsstrahlung 
radiation to produce gamma rays. The latter may produce more pairs and the
pairs produce more gamma rays ... on and on the cascading process goes, until 
ionization becomes the main channel of energy loss for relatively low-energy 
electrons and positrons. Therefore, upon the incident of each gamma ray on 
the atmosphere, a shower of charged particles is formed in the atmosphere. 
Moving downward, the density of shower particles increases until it reaches 
a maximum (known as the shower maximum), typically at an altitude of roughly 
10 km above sea level, and then begins to decrease. The air showers may
be observed through the detection of light emitted (or induced) by the shower particles 
or the detection of those shower particles that manage to reach the ground.

Unfortunately, not all air showers seen are initiated by gamma rays. In fact, 
only a tiny fraction of them are. This is because the showers are also formed 
when cosmic ray particles (mainly protons) interact with the atmosphere, as 
also illustrated in Fig.~\ref{Fig:shower}. Since cosmic rays outnumber cosmic 
TeV gamma rays by many orders of magnitude, picking out gamma-ray-induced 
showers is truly like finding a needle in a haystack! This is a main reason why 
it had taken about two decades of painstaking development before the first 
TeV gamma-ray source, the Crab Nebula, was convincingly detected (Weekes et 
al.~\cite{Weekes89}.) The key for the success lies in the formulation of an 
empirical procedure to separate the showers initiated by gamma rays from 
those by cosmic rays. There are physical differences between electromagnetic 
showers and hadronic showers. Unlike electromagnetic interactions, hadronic 
interactions result mainly in pions. Therefore, hadronic showers mainly 
contain the decaying product of pions, including muons, electrons, positrons, 
and neutrinos from charged pions ($\pi^{\pm}$), as well as gamma rays from 
neutral pions ($\pi^0$). While subsequent $\pi^0$-induced events are 
indistinguishable from the gamma-ray events of interest, $\pi^{\pm}$-induced 
events manifest themselves, e.g., in the associated muons or neutrinos. It 
should be noted that the background events induced by cosmic ray electrons 
are also electromagnetic in origin and are thus difficult to eliminate.  

Two broad classes of experiments are designed to explore the differences between electromagnetic and hadronic showers. One is based on detecting shower particles (photons, muons, electrons, positrons, and neutrinos) that reach the ground, while the other is based on detecting Cherenkov radiation induced by superluminal charged particles in air showers. The main advantages of the former include a long duty cycle and a large field-of-view, while those of 
the latter include high sensitivity (due to efficient gamma--hadron separation), low energy threshold, and good energy and spatial resolution. To a large extent, therefore, the two types of experiments are complementary in practice (e.g., wide-field surveying vs narrow-field imaging). The examples of particle-based experiments include Milagro (which is no longer in operation) and AS$\gamma$ and ARGO, both of which are located at the Yangbajing (YBJ) International Cosmic Ray Observatory in Tibet, China and are ongoing.  Cherenkov experiments can be further divided into imaging experiments, such as CANGAROO-III, HESS, MAGIC, and VERITAS, and non-imaging experiments, such as CELESTE and STACEE. The imaging technique was pioneered by the Whipple Collaboration, which led to the detection of the very first TeV gamma-ray source, and was greatly enhanced by the HEGRA Collaboration through stereoscopic imaging with multiple telescopes. The stereo-imaging techniques is employed in the current (and future) generation of narrow-field imaging experiments. In general, the imaging experiments are far more sensitive than their non-imaging counterparts. The duty cycle of Cherenkov experiments is limited by the requirement of their operating under good weather on moonless nights. For example, VERITAS typically runs for 700--800 hours in a year, or a duty cycle of $<$ 10\%, compared to the $>$ 90\% duty cycle of, e.g., Milagro. More technical details can be found in a recent review article by Aharonian et al. (\cite{Aharev08}).

\subsection{Development and Scientific Drivers}
\label{subsect:development}

The primary driver for the development of TeV gamma-ray astronomy is to 
utilize the unique window of opportunity on the ground to push astronomy 
towards the uppermost end of the electromagnetic spectrum. As the history 
of astronomy has shown, a new window into the universe nearly always brings 
about new discoveries. The prospect of probing the most energetic and most 
violent phenomena in the universe provided strong motivation for decades of 
painstaking efforts to develop and perfect techniques for the field. 

TeV gamma ray astronomy attempts to address many of the same questions 
that other branches of astronomy do. They include: cosmic sources of TeV 
photons, radiation geometries and mechanisms, properties of radiating 
particles and their environments, and so on. The field also offers an 
excellent example of interdisciplinary, collaborative efforts between 
astronomers and physicists, because it also explores topics that go beyond
``traditional'' astronomy, including acceleration of particles in various 
astronomical settings, the origin of cosmic rays, the nature of dark matter,
and cosmology in general.
The interdisciplinary nature of the field is also reflected in the data
collection, reduction, and analysis procedures. As already described in 
\S~\ref{subsect:principles}, the experiments employ instrumentation that 
is familiar both to astronomers (e.g., optical telescopes) and particle
physicists (e.g., detectors). In data reduction, cuts are made to separate 
gamma-ray and cosmic-ray events, which bears resemblance to event selection
in a particle physics experiment. The products of data analysis are, 
however, quite standard in astronomy, such as light curves, spectra, 
and images.

\begin{figure}
   \vspace{2mm}
   \begin{center}
   \hspace{3mm}\psfig{figure=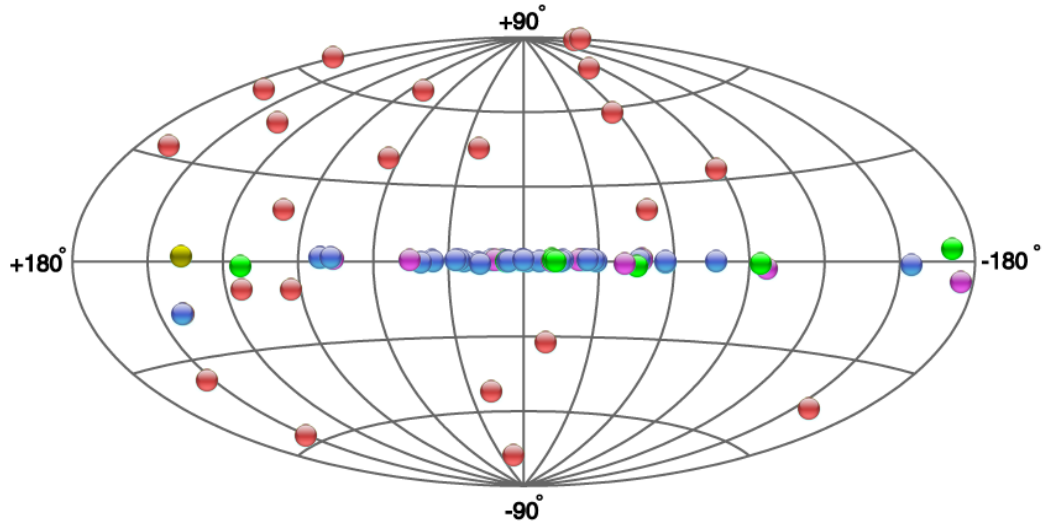,width=140mm,angle=-90.0}
%   \parbox{180mm}{{\vspace{2mm} }}
   \caption{Distribution of discrete TeV gamma-ray sources (as of March 2009).
The sky map is in galactic coordinates. 
The colors differentiate various classes of sources (for legend see 
http://tevcat.uchicago.edu). Courtesy of the TeVCat Team}.
   \label{Fig:sky}
   \end{center}
\end{figure}

Since the detection of the Crab Nebula, TeV gamma-ray astronomy has 
experienced steady, albeit slow at times, growth 
throughout the 1990s and in the early 2000s, and has matured significantly 
over the past five years or so, thanks to the availability of a new 
generation of Cherenkov gamma-ray observatories. The number of sources 
detected has grown rapidly from a handful to over 70 (Aharonian et 
al.~\cite{Aharev08}, and references therein). More significantly, an 
increasing number of classes of sources have been established as TeV 
gamma ray emitters, including BL Lac objects, radio galaxies, quasars, shell-type 
supernova remnants (SNRs), pulsar wind nebulae (PWNe), X-ray binaries, and 
stellar clusters, as shown in Fig.~\ref{Fig:sky}. 
 
Besides discrete sources, large-scale diffuse TeV gamma ray emission has also 
been detected along the ``Galactic Ridge'' (Aharonian et al.~\cite{GC_ridge}) 
and in the Cygnus region (Abdo et al.~\cite{Cygnus}), offering direct 
evidence for interactions between cosmic rays and molecular clouds. The 
latter might be a significant contributor to the reported excess of signals 
in the Cygnus region that could be associated with Galactic cosmic-ray 
particles and gamma rays (Amenomori et al.~\cite{Amenomori06}).

\section{Scientific Achievements}
\label{sect:achievements}

Much has been learned about the TeV gamma-ray sky over the past two decades.
In the following, we will focus on the progress that has been made recently 
on a few selected topics. This article is meant to provide an update on an 
earlier review (Cui~\cite{Cui06}).

\subsection{Origin of Cosmic TeV Gamma Rays}
\label{subsect:origin}

The precise mechanism for producing the detected TeV photons in an 
astronomical environment is still not entirely understood. It most likely 
varies from one class of sources to another, or even from one source to 
another in the same class. The proposed theoretical scenarios fall into two 
broad categories, leptonic models and hadronic models, although hybrid 
scenarios have also been put forth. The models differ mainly in the physical 
nature of gamma-ray emitting particles. Although details vary for different 
classes of sources, the general features are quite similar. 

In leptonic models, 
TeV gamma rays are attributed to the inverse Compton (IC) scattering of 
low-energy photons by relativistic non-thermal electrons (or positrons). 
The sources of seed photons may originate from synchrotron radiation from 
the leptons themselves (i.e., synchrotron self-Compton, or SSC for short), 
or from external radiation fields associated with, e.g., accretion disk, 
disk corona, companion star, broad-line region, or dusty torus, or from 
cosmic background radiation, or some combination of these. In hadronic 
models, the TeV gamma-ray emission is thought to be associated with the 
decaying of $\pi^0$, which may be produced in $pp$ or $p\gamma$ interactions, 
or even with the synchrotron radiation from ultra-relativistic protons in a 
strong magnetic field in some cases. In reality, it is entirely possible 
that both leptonic and hadronic processes are operational, thus giving the hybrid
scenarios, but not necessarily contributing equally to the observed gamma 
ray emission.

A related question is regarding the acceleration of particles in astronomical
environments, which is even less understood. It is often assumed that strong
shocks are responsible, via the first-order Fermi mechanism. This has 
provided a theoretical basis for looking for GeV--TeV gamma rays from sources
that are known to produce strong shocks. The approach is fairly successful in 
practice. The steady-state spectral energy distribution of accelerated 
particles is determined by the balance of timescales related to acceleration, 
heating, and cooling processes. For rapidly varying sources, however, 
non-equilibrium effects may become very important in the study of, e.g.,
spectral variability. Time-dependent calculations must be performed in such
cases.

\subsubsection{Active Galactic Nuclei}
\label{subsubsect:agn}

Active galactic nuclei (AGNs) were among the first sources detected at TeV
energies and have remained the largest source population for TeV
gamma-ray astronomy. Initially, the TeV gamma-ray emitting AGNs found were 
exclusively in BL Lac objects at relatively low redshifts. With lower energy
thresholds and better sensitivities offered by the new generation of ground-based 
gamma-ray observatories, significant progress has been made in recent years 
on three fronts: (1) much improved quality of multiwavelength data on a 
few bright TeV blazars for detailed studies; (2) detection of flat-spectrum 
radio quasars whose gamma-ray output peaks at lower energies than BL Lac 
objects; and (3) detection of AGNs at higher redshifts.

\begin{figure}
   \vspace{2mm}
   \begin{center}
   \hspace{3mm}\psfig{figure=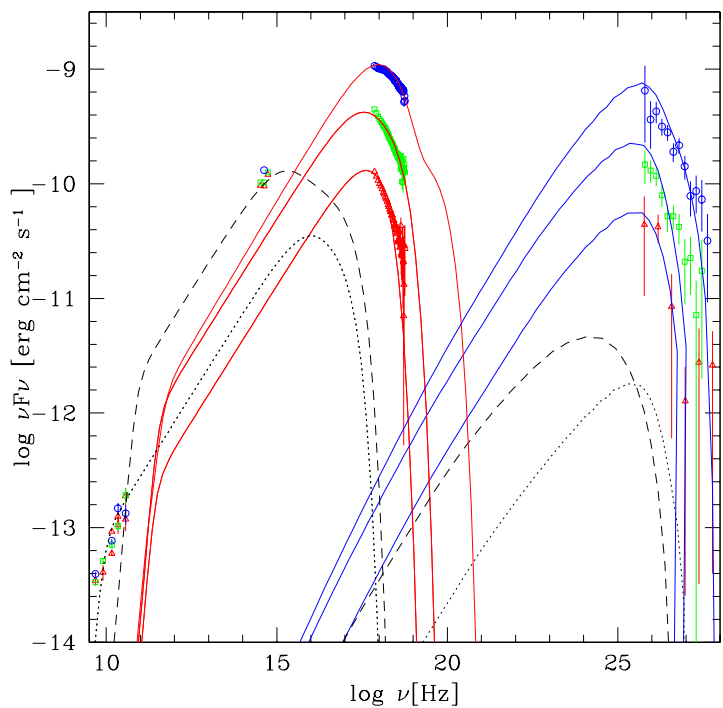,width=100mm,angle=0.0}
%   \parbox{180mm}{{\vspace{2mm} }}
   \caption{Spectral energy distribution of Mrk 421. Three sets of 
flux-averaged SEDs are shown in red, green, and blue, with the flux
increasing roughly by a factor of 3 at each step both at X-ray and
gamma-ray energies. Note that the optical and radio fluxes vary little. 
The lines show the results of a multi-zone model. Adapted 
from B{\l}a\.zejowski et al.~\cite{Blazejow05}}.
   \label{Fig:m4sed}
   \end{center}
\end{figure}

BL Lac objects and flat-spectrum radio quasars both belong to a sub-class 
of AGNs, which are known as blazars. Blazars are known to be highly variable across
nearly the entire electromagnetic spectrum, producing flares of duration 
as short as a few minutes or outbursts of duration as long as many months. 
This, coupled with the 
non-thermal nature of their emission, implies that the photons from blazars
most likely originate in the jets that are directed roughly towards
us (Urry  \& Padovani~\cite{urry95}). Such jet emission, Doppler-boosted
in intensity as well as in energy, dominates over radiation from all other
sources (e.g., accretion flows). Therefore, blazars are excellent 
laboratories for studying physical processes in the jets of AGNs (or perhaps
of black hole systems in general). 
Not all blazars have been established as TeV gamma-ray emitters. Those that 
have invariably show a characteristic double-peaked spectral energy 
distribution (SED), with one of the peaks located at keV energies and the 
other at TeV energies (see Fig.~\ref{Fig:m4sed} for an example). The two 
peaks are seen to shift in a correlated manner during a major flare (or
outburst). Moreover, both the X-ray and gamma-ray spectra of a blazar tend 
to become flatter as the source brightens.

Various models have been proposed to account for the 
broadband SED of TeV gamma-ray blazars. Both leptonic and hadronic models 
attribute the lower-energy SED peak to synchrotron radiation from relativistic 
electrons (and positrons) in the jets, but they differ on the origin of the 
higher-energy peak. 
The leptonic models invoke the IC process to explain the gamma-ray emission 
from blazars. The seed photons may be the synchrotron photons emitted by the
same electrons (e.g., Maraschi et al.~\cite{Maraschi92}; Bloom \& Marscher~\cite{BM96}) or 
originate from sources external to the jet (e.g., Dermer et 
al.~\cite{Dermer92}; Sikora et al.~\cite{Sikora94}). Such models are strongly 
motivated by the apparent correlation between X-ray and TeV variabilities, 
as well as by the shape of the measured SEDs. However, the simple (thus 
commonly used) homogeneous models are increasingly challenged by observations.
For instance, it is now quite clear that one-zone SSC models cannot account
for the measured broadband SEDs of TeV gamma-ray blazars; more ``zones'' are 
almost certainly needed to explain the radio and optical emission (see, e.g., 
Fig.~\ref{Fig:m4sed}). Also, as the quality of multiwavelength
data improves, it has become apparent that the X-ray and gamma-ray correlation
is fairly loose (B{\l}a\.zejowski et al.~\cite{Blazejow05}; Horan et al.~\cite{Horan09}) or even 
absent in some cases (Acciari et al.~\cite{Gall09}). Perhaps, the most serious challenge comes from the
discovery of TeV gamma-ray flares that either have no counterparts at X-ray 
energies or are significantly offset in time from their X-ray counterparts 
(Krawczynski et al.~\cite{Kraw04}; B{\l}a\.zejowski et al.~\cite{Blazejow05}). 
The 
latter are sometimes referred to as ``orphan TeV flares'', even though they 
might not be genuinely orphans (see discussion in B{\l}a\.zejowski et al.~\cite{Blazejow05}). Inhomogeneous models have been proposed 
(e.g., Ghisellini et al.~\cite{Ghisellini05}). It remains to be 
seen whether they can account for all of the observed properties of TeV 
blazars.

Hadronic processes are more complicated. They could involve proton-induced cascade (Mannheim \& Biermann~\cite{Manheim92}) or $pp$ collisions (Dar \& Laor~\cite{Dar97}; Beall \& Bednarek~\cite{Beall99}; Pohl \& Schlickeiser~\cite{Pohl00}), but it all boils down to $\pi^0$ decay for gamma-ray production (with possible contribution from relativistic leptons via the IC process). For TeV gamma-ray blazars, however, it has been argued that $\pi^0$ decay might not important; instead, it is the synchrotron radiation from ultra-relativistic protons in a strong magnetic field that is responsible for the observed TeV photons from blazars (Aharonian~\cite{Ahaproton00}; M\"ucke et al.~\cite{mucke03}). The hadronic models are also capable of describing the broadband SEDs of blazars (Aharonian et al.~\cite{PKS2155}) and, in principle, of explaining the keV-TeV correlation, if an appreciable amount of X-ray emission comes from synchrotron radiation of {\it secondary} leptons (originating in the decay of charged pions). There is quite a lot of flexibility in this case, because the co-accelerated primary electrons might also contribute to the X-ray band in a significant manner. This might explain the scatters in the keV-TeV correlation. However, like the leptonic models, the hadronic models also face the challenges of, e.g., explaining the ``orphan'' TeV gamma-ray flares. 

In an attempt to account for ``orphan'' TeV gamma-ray flares, B\"{o}ttcher (\cite{Boettcher05}) invoked a scenario, in which both relativistic electrons and protons play a significant role in gamma-ray production. In this case, the ``orphan'' TeV gamma-ray flare is caused by relativistic 
{\em protons} in the jets interacting with externally reflected synchrotron photons that are associated with an earlier X-ray/TeV flare, which itself is associated with relativistic {\em electrons} (i.e., synchrotron+SSC) in the jets. The absence of an IC signal expected from the interaction between the relativistic electrons in the jets and the externally reflected photons is attributable to the Klein-Nishina effects. The time interval between the primary X-ray/TeV flare and the ``orphan'' flare in 1ES~1959+650 appears to be consistent with the propagation of synchrotron photons from the primary flare (to some external cloud and back to the jets), with the chosen model parameters.

\begin{figure}
   \vspace{2mm}
   \begin{center}
   \hspace{3mm}\psfig{figure=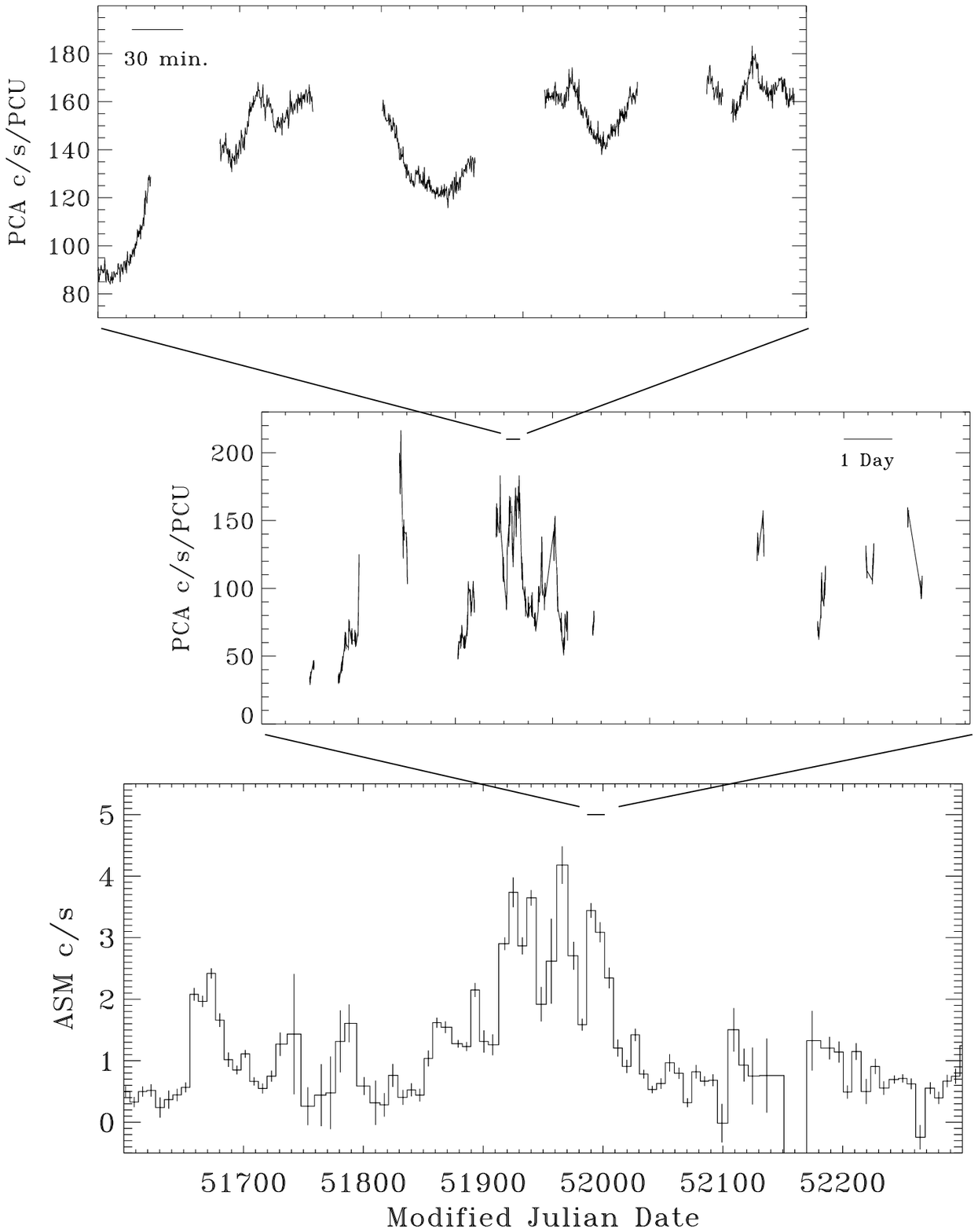,width=73mm,angle=0.0}
   \hspace{3mm}\psfig{figure=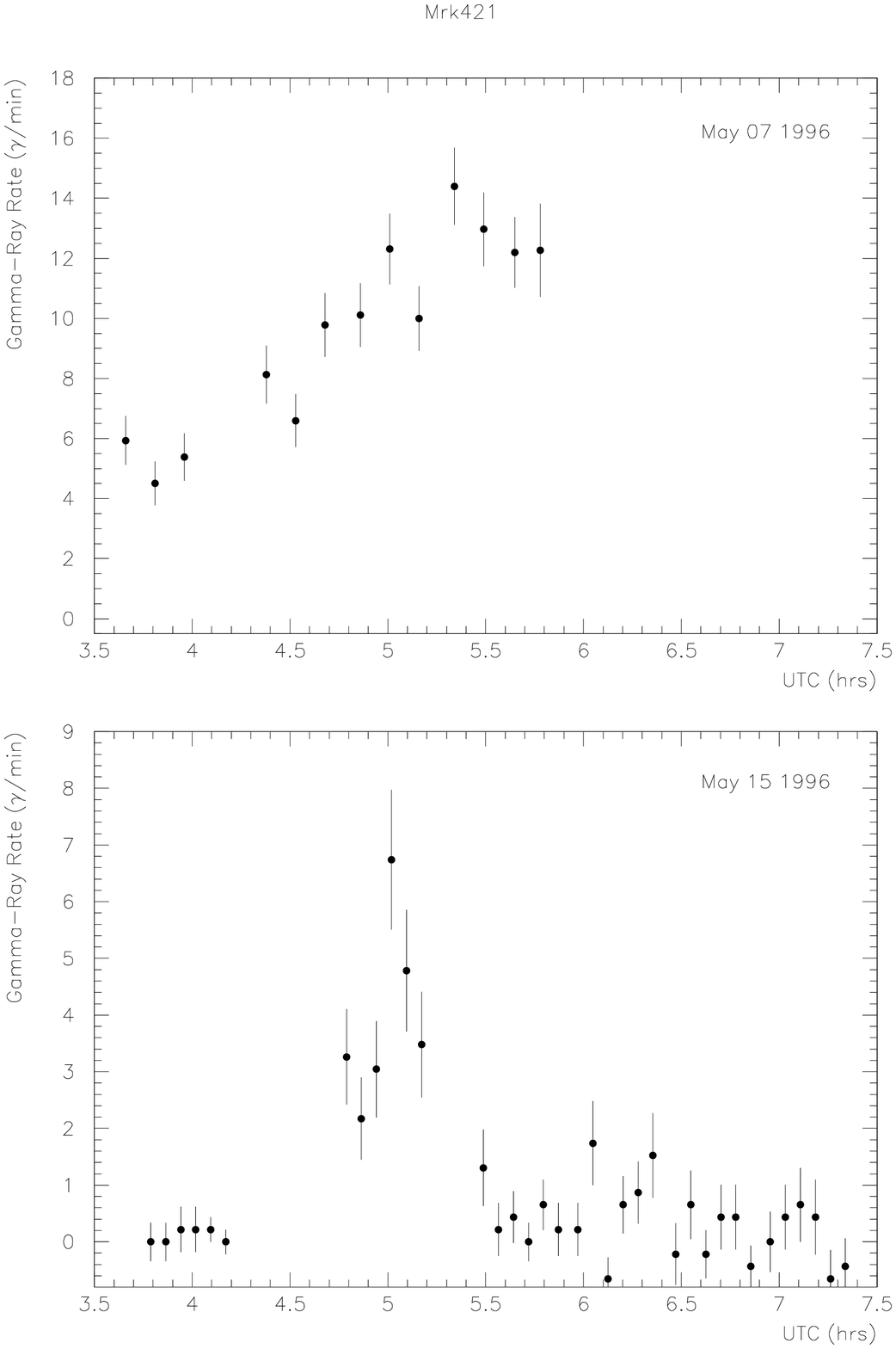,width=68mm,angle=0.0}
%   \parbox{180mm}{{\vspace{2mm} }}
   \caption{Flaring activities of Mrk 421. ({\em left}) X-ray flaring
hierarchy (taken from Cui~\cite{Cui04}); ({\em right}) Rapid TeV flares
(taken from Gaidos et al.~\cite{Gaidos96}). }
   \label{Fig:m4flare}
   \end{center}
\end{figure}

One promising and relatively model-independent approach for distinguishing the models is to study the variability of TeV gamma-ray blazars on very short time scales ($<$ 1 hour). Such rapid variability was first seen in Mrk 421 at TeV energies in the form of a spectacular flare (Gaidos et 
al.~\cite{Gaidos96}). Similar flares have also subsequently been observed at keV energies (Cui~\cite{Cui04}). In fact, the high-quality X-ray data have unveiled a hierarchical nature of the flaring phenomenon. As illustrated in Fig.~\ref{Fig:m4flare}, X-ray flares are seen on a vast range of time scales. Comparing to Mrk 501, it seems that the more frequently flaring occurs on one time scale the more frequently it does on other time scales (Xue \& Cui~\cite{Xue05}). Very recently, TeV gamma-ray variability has been observed on a time scale of minutes in  PKS~2155-304
(Aharonian et al.~\cite{PKSflares}) and Mrk 501 (Albert et al.~\cite{m5flares}). On general physical grounds, these minute-scale TeV gamma-ray flares are very difficult (if at all possible) to be accomodated by the proton-synchrotron mechanism --- an unreasonably large magnetic field would be required to match the synchrotron cooling time of the protons to the observed variability time scale.

The minute-scale gamma-ray flares present a challenge to leptonic models as well. For instance, the observed TeV flare from PKS~2155-304 not only implies a highly compact emission region (comparable to the Schwarzschild radius of a $10^9$ $M_{\odot}$ black hole!) but imposes a {\it lower} limit on the Doppler factor of bulk motion ($>$ 100), which is uncomfortably
large (based on distributions derived from radio observations). The problem may be alleviated by introducing additional features to the models, such as radiative deceleration of jets (Levinson~\cite{Levinson07}). On the other hand, a large Doppler factor would argue in favor of external IC processes (Begelman et al.~\cite{Begelman08}). The added complexity offers more ``wiggle room'' in applying the models to the data. Simultaneous SED modeling helps in providing additional constraints but much degeneracy tends to remain. The situation can be improved by having a sufficiently large sample of rapid flares seen both at keV and TeV energies. This has remained an observational challenge, because it requires dedicated monitoring at the two wavebands, as well as a lot of good luck!

On the theoretical front, the precise mechanism that causes flaring in blazars is not yet understood. The flares could be ultimately related to internal shocks in the jets (Rees~\cite{Rees78}; Spada et al.~\cite{Spada01}), or to major ejection of new components of relativistic plasma into the jet (e.g., B\"ottcher et al.~\cite{Boettcher97}; Mastichiadis \& Kirk~\cite{MK97}),
or to magnetic reconnection events (like solar flares; Lyutikov~\cite{Lyutikov03}). The amplitude and duration of flares probably reflect the energetics and physical timescales involved in the processes. The SED of a TeV gamma-ray blazar is known to evolve significantly during a 
major outburst (or flare with a duration of weeks to months). Both the X-ray and gamma-ray spectra tend to flatten as the source brightens (e.g., Xue et al.~\cite{XYC06}; Krennrich et al.~\cite{Krennrich02}; see, however, Acciari et al.~\cite{Gall09}). It has been shown that the 
observed X-ray spectral variability requires a change in multiple key physical parameters (Xue et al.~\cite{XYC06}), including the total energy, spectral index, and maximum Lorentz factor of synchrotron emitting electrons, as well as magnetic field, as opposed to any single parameter (as argued by Konopelko et al.~\cite{Konopelko03}).

At present, the proposed models for blazars are still at a stage of {\em assuming} a spectral energy distribution for radiating particles. No detailed treatment of particle acceleration is usually made. It is a common practice to adopt a simple power law for the particle distribution, although more sophisticated treatment, taking into account the balance among acceleration, heating, and cooling of the particles, leads to more complicated distributions. In principle, the observed gamma-ray spectrum of blazars imposes constraints on the particle acceleration process in either leptonic or hadronic scenarios. To this end, a recent surprise is the unusually hard {\em intrinsic} gamma-ray spectrum of blazars at moderate redshifts. To obtain the intrinsic spectrum, one must correct for the effects of gamma-ray absorption due to diffuse infrared background radiation (see \S~\ref{subsect:cosmology}). Even for minimum possible absorption, the intrinsic gamma-ray spectrum of, e.g., 1ES 1101-232, is found to be extremely hard (Aharonian et al.~\cite{E1101}). This may have a profound impact on our understanding of particle acceleration in the jets of AGNs.

Until very recently, BL Lac objects represented the only type of AGNs that are 
seen to emit TeV gamma rays. With the detection of 3C~279 (Errando et al.~\cite{3c279}), flat-spectrum radio quasars have now emerged as a viable
source population for TeV gamma-ray astronomy. Equally significant is the
discovery of BL Lac objects, such as W~Comae and 3C~66A, and  (Acciari et al.~\cite{wcomae,3C66A}), whose SED peaks are located in 
between the traditional TeV BL Lac objects and flat-spectrum radio quasars, 
as the observation begins to cover the entire continuum of the blazar 
phenomenon.

The only AGN that have been detected at TeV energies and are not a blazar are
M~87 (Aharonian et al.~\cite{M87hegra}; Acciari et al.~\cite{M87}; 
Albert et al.~\cite{M87magic}) and Cen~A (Aharonian et al.~\cite{cena}). Both sources are 
classified as an FR~I radio galaxy. For 
quite some time, M~87 is thought to be a source of ultra-high-energy particles 
(Ginsburg \& Syrovatskii~\cite{Ginsburg64}. In this sense, the detection 
of M~87 at TeV energies is not a total surprise. As in the case of blazars, 
the TeV gamma-ray emission from M~87 has also been modeled both in the 
leptonic and hadronic scenarios. In the leptonic model, M~87 is simply 
viewed as a ``mis-aligned blazar'' (Bai \& Lee~\cite{Bai01}), with the TeV 
gamma rays originating in the IC process.  

More recently, M~87 has been seen to vary on a timescale of days at TeV energies 
(Aharonian et al.~\cite{M87var}), implying a very compact region of gamma 
ray production 
in the immediate vicinity of the black hole. The previously favored HST-1 
hot spot (e.g., Stawarz et al.~\cite{Stawarz06}) appears to have been ruled
out by the observed gamma-ray variability. The new variability constraints
have prompted the development of more sophisticated leptonic models (Lenain 
et al.~\cite{Lenain08}; Tavecchio \& Ghisellini\~cite{Tavecchio08}). 
Alternatively, the hadronic models attribute the gamma rays to synchrotron 
radiation from protons (Protheroe et al.~\cite{Protheroe02}). Such models 
have not been scrutinized with observations as much as the leptonic models. 

In many ways, Cen~A is similar to M~87, so the same physical processes might 
also operate in this source. Being the nearest active galaxy known, its inner
jets structures could be resolvable with the next-generation gamma ray 
experiments and might thus offer an excellent laboratory for studying 
particle acceleration in the immediate vicinity of black holes.

\subsubsection{Pulsars, Pulsar Wind Nebulae, and Supernova Remnants}
\label{subsubsect:snr}

The Crab Nebula is the first pulsar wind nebula (PWN) detected at TeV energies. The 
TeV gamma-ray emission is still spatially unresolved, although the bulk of 
the emission is clearly associated with the pulsar wind. Only very 
recently, the MAGIC collaboration reported the detection of gamma rays from 
the pulsar itself, with the help of a special triggering scheme (Aliu et al.~\cite{crabpulsar}). 
This is a highly significant result, because it may
shed much light on the issue of gamma-ray production in pulsars, which has 
been a long, intense debate since the detection of pulsars by EGRET. 

Pulsars have long been suspected to be a source of Galactic cosmic-ray 
electrons, so seeing gamma rays from them is not a surprise in itself. The 
prospect of using the gamma rays to study particle acceleration and radiative 
processes in a neutron star environment is, on the other hand, very exciting. 
Now that the first results from {\em Fermi} on pulsars seem to rule out an 
origin of gamma rays near the polar cap of neutron stars, at least for some
pulsars (Abdo et al.~\cite{Fermi_vela}), there is more hope to see additional 
pulsars at TeV energies. 

\begin{figure}
   \vspace{2mm}
   \begin{center}
   \hspace{3mm}\psfig{figure=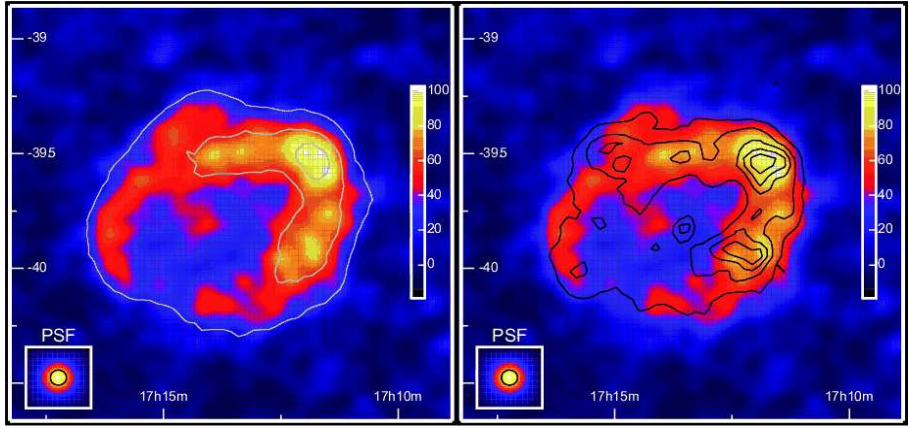,width=120mm,angle=0.0}
%   \parbox{180mm}{{\vspace{2mm} }}
   \caption{TeV gamma-ray image of the supernova remnant RX~J1713.7-3946. The 
superimposed contours show the significance ({\em left}) and 1--3 keV X-ray surface brightness ({\em right}), respectively. Taken from Aharonian et al.~\cite{J1713b}. }
   \label{Fig:rxj1713}
   \end{center}
\end{figure}

The first source showing extended TeV gamma-ray emission is RX~J1713.7-3946 
(Aharonian et al.~\cite{J1713}), a shell-type supernova remnant (SNR). The 
exquisite gamma-ray image obtained of the source (as shown in 
Fig.~\ref{Fig:rxj1713}) has, in our view, truly elevated the field to a new 
level. For the very first time, imaging is carried out at a resolution of 
arcminutes at TeV energies. Not only has this enabled morphological studies 
of extended gamma ray emission over multiple wavebands, it is also critical 
for cross-band identification of new source populations. To put things in 
prospective, the nature of gamma ray bursts (GRBs) had remained elusive until 
imaging of their X-ray afterglows reached a resolution of arcminutes with 
{\em BeppoSAX}. The number of TeV gamma ray emitting PWNe and SNRs has been 
increasing rapidly in over the past five years. They are the dominant 
population among Galactic TeV gamma-ray sources.

The discovery of TeV gamma-ray emission from SNRs has brought a sigh of relief for many in the field, because it is almost taken for granted that they are the main source of Galactic cosmic rays. However, direct evidence for relating the observed gamma rays to hadronic processes is still lacking, because the observed gamma-ray emission can be accommodated by both leptonic and hadronic scenarios (not without issues in either case). In leptonic models, TeV photons are thought to be produced in the IC scattering of mainly CMB photons by relativistic electrons that are accelerated by a strong shock at the outer edge of an SNR (e.g., Aharonian et al.~\cite{J1713a,G0.9}; Brogan et al.~\cite{G12.82}),  although other sources of seed photons (e.g., star light) may be needed in some cases to account for the observed broadband SED. For RX~J1713.7-3946, the leptonic models can naturally explain the observed spatial coincidence between the X-ray and TeV emitting regions, since X-ray photons are attributed to synchrotron radiation from the same electrons. Quantitatively, however, a low magnetic field ($B < 15$ 
$\mu G$) in the shell would be required to account for the observed SED (Aharonian et al.~\cite{J1713a}), which appears to be at odds with the observed X-ray variability associated with the shell on a timescale of about a year (Uchiyama et al.~\cite{Uchiyama07}).

On the other hand, the observed TeV emission from SNRs can also be explained as the product of the decay of neutral pions that are produced in the collision between the relativistic protons and surrounding medium (Aharonian et al.~\cite{J1713a}). In this case, the challenge is to account for the observed X-ray/TeV spatial correlation. One can either attribute X-rays to synchrotron radiation from {\it co-accelerated} electrons or invoke correlated enhancement of the magnetic field and the density of the surrounding medium (Aharonian et al.~\cite{J1713a}). At present, it is fair to say that no conclusive evidence exists for the acceleration of protons in SNRs. Somewhat disturbingly, this is not due to insufficient quality of the gamma-ray data. For instance, the gamma ray spectrum of RX~J1713.7-3946 is of high quality over three decades in energy, from about 0.1 TeV to nearly 100 TeV (Aharonian et al.~\cite{J1713b})! In this particular case, however, the multiwavelength observations seems to favor a hadronic origin (Tanaka et al.~\cite{Tanaka08}), because it provides a more reasonable fit to the SED and can also explain the year-scale X-ray variability. 
 
\begin{figure}
   \vspace{2mm}
   \begin{center}
   \hspace{3mm}\psfig{figure=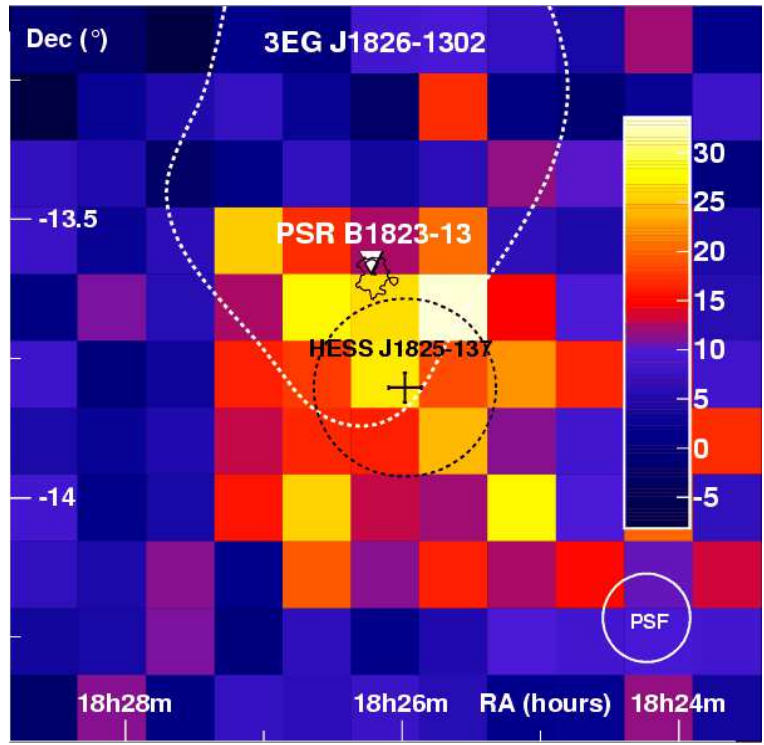,width=100mm,angle=0.0}
%   \parbox{180mm}{{\vspace{2mm} }}
   \caption{TeV gamma-ray image of HESS J1825-137. The cross shows the 
best-fit centroid of the gamma-ray excess, while the black dotted circle 
shows the best-fit 1$\sigma$ emission region size assuming a Gaussian 
brightness profile. The position of PSR~B1823-13 is indicated by the white 
triangle. The black contours denote the X-ray emission. The 95% confidence 
region (dotted white line) for the position of the unidentified EGRET source 
3EG J1826-1302 is also shown. Taken from Aharonian et al.~\cite{G18}.}
   \label{Fig:j1825}
   \end{center}
\end{figure}

For PWNe, leptonic scenarios are more likely, as supported by similar morphologies at keV and TeV energies. The X-ray emission is associated with synchrotron radiation from relativistic electrons, while TeV gamma rays are produced via IC processes, with seed photons coming mainly from the CMB (e.g., Aharonian et al.~\cite{MSH,G18,VelaX,G313}). Intriguingly, however, 
there are also cases in which there is little morphological correspondence between TeV gamma-ray and X-ray images. Fig.~\ref{Fig:j1825} shows such an example. The connection between HESS~J1825-137 and the PWN associated with PSR~B1823-13 was made based on similar asymmetric distributions at TeV and keV energies roughly along the north-south direction (Aharonian et al.~\cite{G18}). However, the gamma-ray emission is much more extended than 
the X-ray emission, and the two are significantly offset from each other. This has added uncertainty to the identification of a few TeV gamma ray sources that are in the vicinity of pulsars (and might thus been associated with PWNe; e.g., Cui \& Konopelko~\cite{CK07}; Chang et al.~\cite{CKC08}). Plausible causes for the morphological differences have been suggested (Aharonian et al.~\cite{G18}). We think that the challenge is to explain, in a natural way, why the processes (e.g., diffusion beyond the PWN boundary) are not operating in systems in which the X-ray and gamma-ray images match well.

\subsubsection{X-ray Binaries}
\label{subsubsect:xrb}

From a historical perspective, X-ray binaries have played a crucial role in 
the development of TeV gamma-ray astronomy. They are among the very first 
sources that were claimed to be TeV gamma-ray emitters in the 70s and 80s. 
The chain reaction fueled by such claims helped generate and sustain the 
momentum to develop more sophisticated experiments and data analysis 
techniques, which ultimately led to the establishment and success of an 
exciting field. It is, however, ironic that none of the early claimed 
detections of X-ray binaries is now deemed credible, because the sources 
have not been seen with the new and much more sensitive experiments over 
the past two decades or so. While one could always invoke once-in-a-lifetime 
transient phenomena to explain away the modern non-detections, it would seem 
to be too much of a coincidence for the source population as a whole to 
cooperate in such a way!

It was not until several years ago that the first credible detection of an 
X-ray binary (PSR~B1259-63) was reported (Aharonian et al.~\cite{B1259}). 
The source consists of a 48-ms radio pulsar and a Be star in a highly 
eccentric orbit (with a period of $\sim$3.4 yrs). Be stars are known to be 
fast rotators that produce a dense equatorial wind. When the neutron star 
passes through the wind, enhanced accretion onto the neutron star (due to 
the capture of wind) is thought to be responsible for the activities 
previously seen in the X-ray and soft gamma-ray bands . Moreover, the 
collision between pulsar wind and stellar wind could result in the formation 
of a strong shock, which is not fundamentally different from the formation of PWN, 
although the wind dynamics in a binary system is certainly different from 
those around an isolated neutron star. 

\begin{figure}
   \vspace{2mm}
   \begin{center}
   \hspace{3mm}\psfig{figure=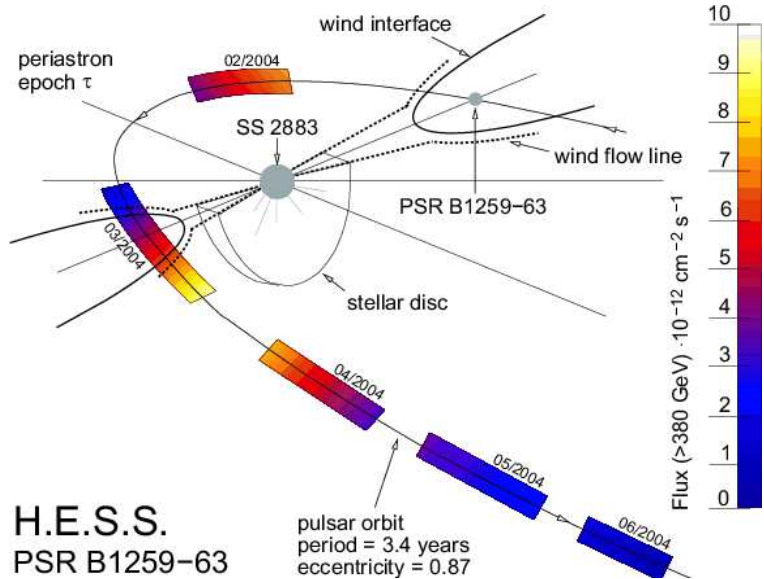,width=100mm,angle=0.0}
%   \parbox{180mm}{{\vspace{2mm} }}
   \caption{Variable TeV gamma-ray emission from PSR B1259-63 over the 
binary orbit. Taken from Aharonian et al.~\cite{B1259}.}
   \label{Fig:psrb1259}
   \end{center}
\end{figure}

The observed TeV gamma rays may originate in the relativistic electrons 
that are accelerated by the shock via the IC process. In this case, however, 
the seed photons are likely dominated by radiation from the Be star. Given 
that, one would expect strong orbit modulation of the TeV gamma-ray emission, 
which appears to be present. Also, the TeV gamma-ray emission is seen to vary 
significantly over the orbit, as shown in Fig.~\ref{Fig:psrb1259}.  It is 
interesting to note that the TeV emission appears to be at a lull at the 
periastron passage. In general, the observed light curve is not compatible 
with simple IC scenarios (Aharonian et al.~\cite{B1259}). No detailed
hadronic modeling has been performed. Given the presence of a dense wind 
disk around the Be star, the $pp$ process might be quite efficient here, 
producing neutral pions, which then decay to produce the detected gamma rays (Kawachi et al.~\cite{Kawachi04}).

During its survey of the Galactic central region, HESS discovered another
X-ray binary, LS~5039 (Aharonian et al.~\cite{LS}). The source stood out
as  the only point-like source detected in the survey. It is only about 
1$^{\circ}$ away from HESS~J1825-137 (a likely PWN), which again illustrates 
the necessity of high-resolution imaging for such a discovery. LS~5039 was 
originally identified by Motch et al. (\cite{Motch97}) using ROSAT data as a 
likely High Mass X-ray Binary (HMXB) system, with an O7V(f) luminous 
companion at a distance of 3~kpc. The nonthermal synchrotron radio emission 
was later discovered by Mart\'{\i} et al. (\cite{Marti98}) using the Very 
Large Array (VLA). The RXTE observations performed by Rib\'o et al. 
(\cite{Ribo99}) show a hard X-ray spectrum extending up to 30~keV, which 
can be fitted satisfactorily with a power-law (plus a strong iron line 
centered at 6.6~keV). Radio interferometric observations with the Very 
Long Baseline Array (VLBA) by Paredes et al. (\cite{Paredes00}) resolved
the source into milliarcsecond bipolar radio jets, suggesting that LS~5039
might be a microquasar. This is supported by dynamical measures that seem 
to favor the presence of a black hole of 3.7M$_{\odot}$ in the system 
(Casares et al.~\cite{Casares05b}), although the uncertainty is still quite 
large.

\begin{figure}
   \vspace{2mm}
   \begin{center}
   \hspace{3mm}\psfig{figure=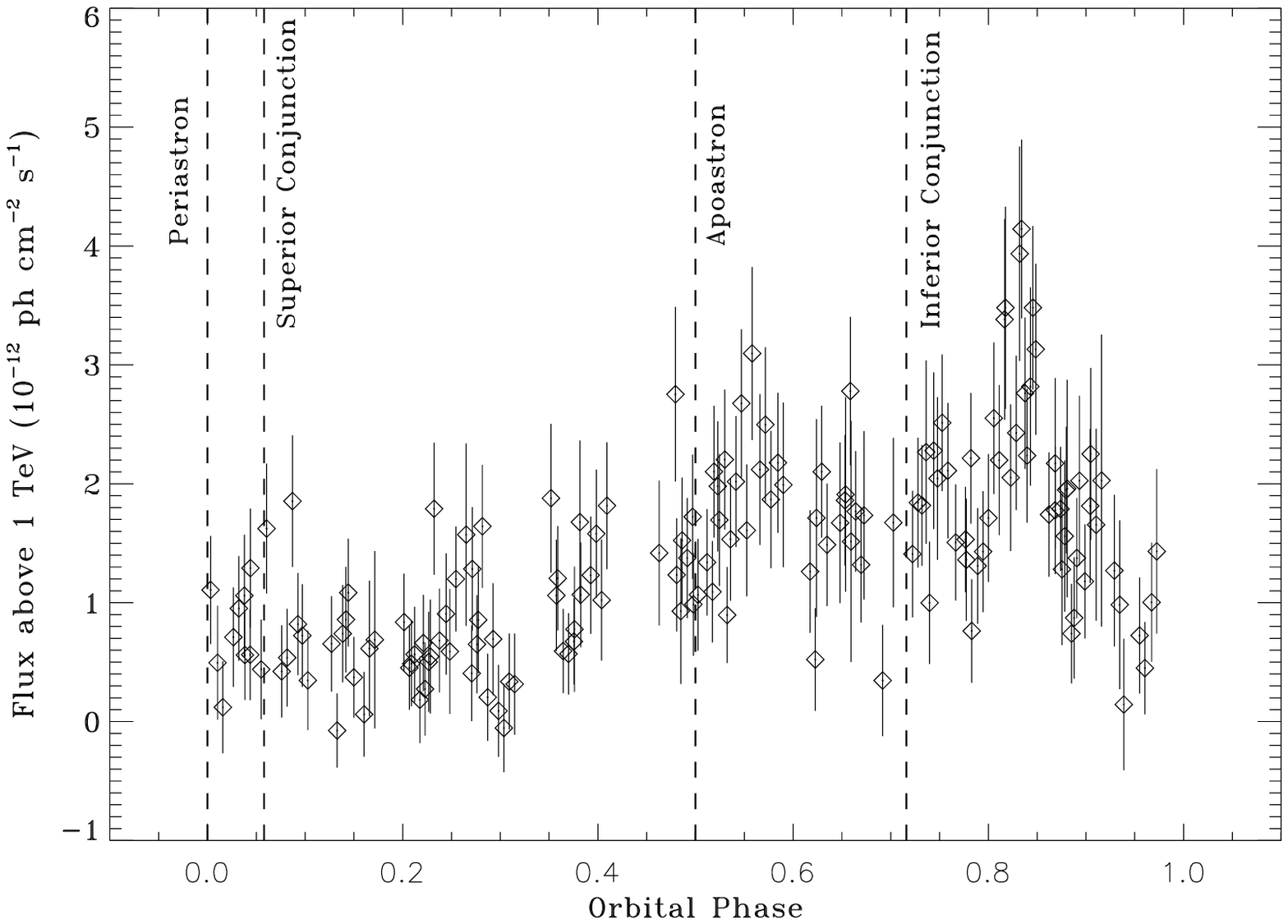,width=70mm,angle=0}
   \hspace{3mm}\psfig{figure=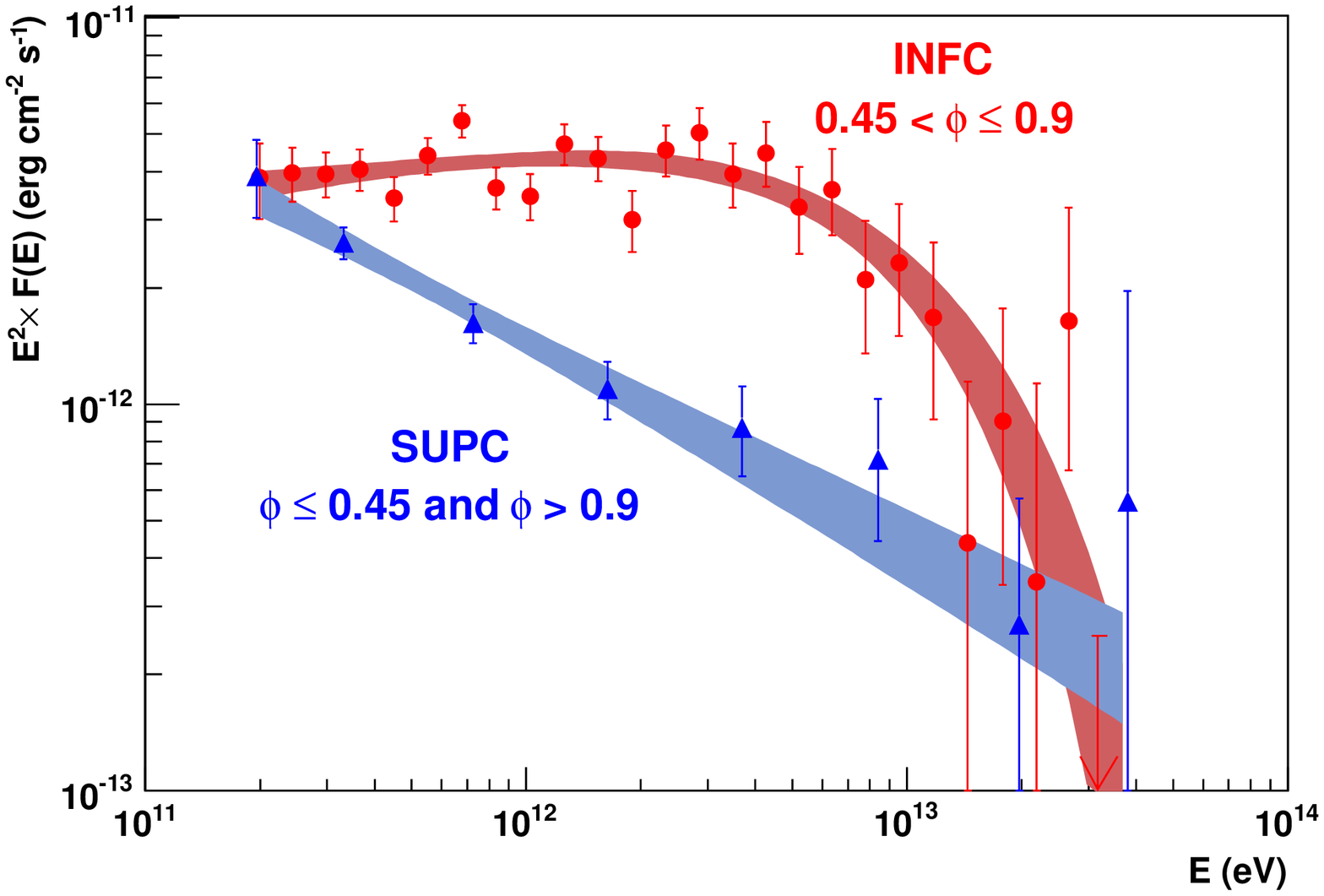,width=70mm,angle=0}
%   \parbox{180mm}{{\vspace{2mm} }}
   \caption{Variable TeV gamma-ray emission from LS 5039 over the 
binary orbit: ({\em left}) folded light curve and ({\em right}) gamma-ray
spectra. Taken from Aharonian et al.~\cite{LSorb}.}
   \label{Fig:ls5039}
   \end{center}
\end{figure}

The orbital period of LS~5039 is only about 3.9 days, so it is relatively easy to quantify the orbital effects on gamma-ray production (compared to PSR~B1259-63). As shown in Fig.~\ref{Fig:ls5039}, not only has the modulation of the gamma-ray flux been well established along the binary orbit, the observed gamma-ray spectrum also shows a very intriguing pattern of variability (Aharonian et al.~\cite{LSorb}). The pattern is not easy to understand in a natural way; 
it might simply reflect, at least partially, a change in the spectral energy distribution of radiating particles along the orbit (e.g., Sierpowska-Bartosik \& Torres~\cite{SierTorres09b}. Of course, other effects are also expected to play a significant role in causing the orbital  modulation of gamma rays, including anisotropic IC scattering, attenuation due to $\gamma \gamma$ 
pair production and subsequent electromagnetic cascading, and adiabatic losses (e.g., Khangulyan et al.~\cite{khangulyan08}).

The enigmatic gamma-ray source LS~I~+61 303 has also recently been seen to
emit TeV gamma rays (Albert et al.~\cite{LSI}; Acciari et al.~\cite{LSIver}). 
Like LS~5039 and PSR~B1259--63, it falls in the category of high-mass X-ray 
binaries (HXMBs). The nature of the compact object is even less certain in 
this case than LS~5039. It was initially
argued to be a neutron star although a more recent study has shown that it 
could also be a black hole, given the uncertainty in the inclination of the 
binary system (Casares et al.~\cite{Casares05a}). On the other hand, the
binary parameters are well determined. The companion star is a Be star, as
in the case of PSR~B1259--63. It is in an eccentric orbit ($e \approx 0.7$) 
with the compact object, with an orbital period of 26.4960 days, which is
determined from  the observed periodic modulation of the radio emission 
(Gregory~\cite{gregory02}). For reference, the periastron passage occurs 
at Phase 0.23. 

For a long time, LS~I~+61~303 is thought to be the counterpart of 2CG 135+01, 
a COS-B source, and more recently that of 3EG J0241+6103, an EGRET 
unidentified source. If the latter association is real, the EGRET 
observations would indicate that the source also varies significantly over 
the binary orbit also at GeV energies. Significant variability is now well
established at TeV energies (Albert et al.~\cite{LSI}; Acciari et al.~\cite{LSIver}), although it is not entirely clear whether it is entirely
related to the orbital motion (however, see Albert et al.~\cite{LSIorb}).
What is most surprising is the fact that the TeV gamma-ray flux peaks near
the {\em apoastron} passage (i.e., when the compact object is farthest from 
the Be star), because, at longer wavelengths, Be binaries tend to become
active near the {\em periastron} passage, presumably due to enhanced 
accretion from the stellar wind. In principle, the gamma-ray lull near the
periastron passage could be the manifestation of $\gamma \gamma$ attenuation
(in the intense stellar radiation field), even though gamma-ray production
is expected to be high as well (either via the IC process in leptonic
scenarios or $pp$ collisions in hadronic scenarios). Calculations have shown
that, with the right choice of parameters, the observed variability can be
accounted for (Sierpowska-Bartosik \& Torres~\cite{SierTorres09a}).

It is worth noting that the three X-ray binaries detected at TeV energies
are all HMXBs. Observationally, accreting X-ray pulsars tend to be found in 
HMXBs, especially in Be systems. Whether there is a causal connection 
between the two remains to be seen. The interaction between pulsar wind and
stellar wind certainly provides a plausible mechanism for accelerating particles
to high energies, which then radiate TeV gamma rays. On the other hand, 
microquasars are predominantly low-mass X-ray binaries (LMXBs), including
some of the most spectacular systems, such as GRS~191+105, that give rise
to the name ``microquasar''. From a theoretical point of view, the jets
in microquasars are certainly promising sites for particle acceleration and
gamma-ray production. The presence of non-thermal particles in the jets is
well established by observations at longer wavelengths (radio and X-ray in
particular). However, no LMXB has been seen at TeV energies (or GeV energies).
This could be due to the lack of (stellar) seed photons for the IC process
in leptonic scenarios, or the lack of strong stellar wind for $pp$ 
collisions in hadronic scenarios, or the lack of particles at sufficiently
high energies. The observations appear to disfavor the SSC process, because
the jets seem more energetic and prominent in LMXBs than in HXMBs.

To make progress, a systematic observational effort is required not only to 
increase the number of TeV gamma-ray emitting X-ray binaries but also to collect 
{\it simultaneous} multiwavelength data. A reliable SED would establish 
at which wavelengths a source radiates most of its power, which is an 
important step towards understanding radiation mechanisms. To date, modeling 
efforts have mainly been based on multiwavelength data taken at different
times. Given that X-ray binaries are known to vary on a wide ranges of 
timescales, the results should be taken with a grain of salt.

\subsubsection{Unidentified TeV Gamma-Ray Sources}
\label{subsubsect:unids}

Arguably the most intriguing recent development in the field is the 
discovery of a population of TeV gamma-ray sources that appear to have no 
counterparts at longer wavelengths. Of course, {\it unidentified} does 
not necessarily imply {\it unidentifiable} or the lack of plausible 
counterparts. The presently unidentified TeV gamma-ray sources cluster 
around the Galactic plane, indicating that they are likely of Galactic 
origin. Moreover, most of them are spatially extended, suggesting that 
they might be associated with unseen PWNe or SNRs, given PWNe and SNRs 
constitute the largest populations of Galactic TeV gamma ray sources. 
Indeed, a number of previously unidentified gamma ray sources have 
subsequently been found to be probably associated with SNRs or PWNe (e.g., 
Aharonian et al.~\cite{J1825,G12.82}; Cui \& Konopelko~\cite{CK07}; 
Tian et al.~\cite{J1731}; Chang et al.~\cite{CKC08}).

TeV~J2032+4130 is the first {\em bona fide} unidentified TeV gamma-ray source 
and has remained as such. It was accidentally discovered by the HEGRA 
Collaboration (Aharonian et al.~\cite{J2032}), in an intense campaign which studied
Cygnus X-3. The initial detection 
was only of marginal statistical significance but has since been confirmed 
by multiple experiments (Aharonian et al.~\cite{HESS2032}; Konopelko et al.~\cite{VER2032}; Albert et al.~\cite{MAGIC2032}). The 
HEGRA results indicated that TeV~J2032+4130 was an extended TeV gamma-ray
source, with a Gaussian radius of 6.2' $\pm$1.2' $_{stat}$ $\pm$
0.9' $_{sys}$, and the center-of-gravity (CoG) of the gamma-ray 
emission lay roughly 0.5$^{\circ}$ north of Cygnus X-3. The measured 
gamma-ray spectrum was hard, with a power-law photon index of -1.9. These 
results have since been confirmed by the MAGIC measurements. 

The situation is complicated by a reported detection of TeV~J2032+4130 at a 
significantly higher flux by the Whipple Collaboration (Lang et al. 2004), 
based on archival observations. The difference in flux could be explained 
by variability of the source over a time scale of years but it would seem to
be at odds with the extended nature of the TeV gamma-ray emission. One 
could speculate about the presence of another variable gamma-ray source that is 
located very close to TeV~J2032+4130. It is worth noting that the peak of 
the gamma-ray emission detected with {\em Whipple} appears to be offset 
by that from the HEGRA position by about 3.6', although the uncertainty is quite 
large. The gamma-ray flux derived from more recent Whipple observations of 
TeV~J2032+4130 is much closer to the HEGRA flux (Konopelko et al.~\cite{VER2032}).

TeV~J2032+4130 lies in the general direction of Cygnus OB2, a rich cluster
of OB stars less than 2 kpc away. For a long time, it has been thought that the 
winds of massive stars in such a cluster carry a sufficient amount of energy 
that, when released, may power the production of very-high-energy (VHE) 
gamma rays through the production of neutral pions in the collisions 
between non-thermal ions (accelerated by the shocks in the winds) and thermal 
protons in the winds (White \& Chen~\cite{WC92}; Torres et al.~\cite{TDR04}). 
Even among the most massive star clusters, Cygnus OB2 represents an extreme case --- it is 
the most massive stellar association known in the Galaxy, containing about 
2600 OB stars (Kn\"{o}dlselder~\cite{knodlseder00}). It is, therefore, natural 
to speculate about a plausible physical connection between TeV~J2032+4130 and Cygnus OB2 (Torres 
et al.~\cite{TDR04}). However, Cygnus OB2 represents quite a large region in 
the sky.

Very recently, a catalog of bright gamma-ray sources has been released that 
contain all sources detected at a statistical significance of $>$ 10$\sigma$ 
with the Large Area Telescope (LAT) on board the {\em Fermi Gamma-ray Space 
Telescope} (Abdo et al.~\cite{Fermi_cat}). In it, 0FGL~J2032.2+4122 lies only 
about 8\arcmin\ from the CoG of TeV~J2032+4130. Note that the 95\% error radius of the LAT position of the source is determined to be 5.1\arcmin and that an overall uncertainty on the position of TeV~J2032+4130 is about 3 \arcmin\ . Therefore,  0FGL~J2032.2+4122 is a 
promising candidate for being the GeV counterpart of TeV~J2032+4130, based on spatial 
coincidence alone.

Interestingly, 0FGL~J2032.2+4122 is one of the 29 gamma-ray pulsars detected 
by {\em Fermi} LAT (Abdo et al.~\cite{Fermi_cat}). It is designated as 
LAT PSR J2032+41, because it falls in a special category of pulsars which seem to 
only show pulsation in the gamma-ray band. The discovery of such pulsars is a 
major highlight at the early stage of the {\em Fermi} mission. Being ``dark'' 
at longer wavelengths, these sources could have easily escaped previous pulsar 
searches or surveys. They also provide a natural explanation for some of the 
unidentified gamma-ray sources, such as TeV~J2032+4130, now that PWNe 
constitute 
a major population among VHE gamma-ray emitters. Additional support for the PWN
nature of TeV~J2032+4130 comes from the detection of extended X-ray emission that 
spatially coincides with TeV~J2032+4130, based on a deep exposure of the region 
with the {\em XMM-Newton} (Horns et al.~\cite{Horns07}). The X-ray feature is said
to be similar in extent to TeV~J2032+4130. The X-ray spectrum is also very hard, 
with a power-law photon index of -1.5. The overall X-ray power output is 
comparable to the TeV gamma-ray power of TeV~J2032+4130.

The significance of establishing a connection between TeV~J2032+4130 and 
LAT~PSR~J2032+41 goes far beyond identifying an unidentified gamma-ray 
source. The  gamma-ray-only pulsars discovered by {\em Fermi} may represent the tip of the
iceberg of a population of such sources. This would imply that other 
unidentified
TeV gamma-ray sources might simply be PWNe associated with these ``dark'' 
pulsars and thus provide the answer to a long-standing question on the nature 
of unidentified gamma-ray sources (from the EGRET days to the TeV gamma-ray era). 
The fact that nearly all unidentified gamma-ray sources are extended is 
consistent with such a scenario. A systematic investigation on the subject, 
utilizing data both from {\em Fermi} and ground-based TeV gamma-ray 
observatories, 
may prove fruitful in unveiling such connections.

One complication in identifying the unidentified TeV gamma-ray sources arises
from the fact that the morphology of the sources may look different at
longer wavelengths. This is perhaps best illustrated by the association of
HESS~J1825-137 with the PWN G18.0-0.7 (Aharonian et al.~\cite{J1825}). 
The gamma-ray emission spreads over a much larger region than the X-ray
emission from the PWN (see Fig.~\ref{Fig:j1825}). Moreover, the pulsar 
(PSR~B1823-13) is offset from
the center of the PWN (which in turn is offset from the CoG of the 
gamma-ray emission). It is the similar asymmetry in the X-ray and gamma-ray
emission with respect to the pulsar that provides additional evidence for
a PWN origin of HESS~J1825-137. Cases like this are not uncommon (e.g., 
Cui \& Konopelko~\cite{CK07}; Chang et al.~\cite{CKC08}), which
makes arguments based on spatial coincidence alone suspect. This is certainly
an area to which improved sensitivity and angular resolution can add much.

\subsection{Impact on Cosmology}
\label{subsect:cosmology}

TeV photons may interact with infrared photons to produce electron-positron 
pairs and thus be effectively ``absorbed''. This process must be taken into
account in modeling the SED of all TeV gamma-ray sources. Cosmologically, the 
implication is that the visible TeV gamma-ray sky does not extend very far, 
due to the presence of permeating infrared background radiation. On the other
hand, we could use distant TeV gamma-ray sources as cosmic beacons to probe
the diffuse infrared background, which has remained an observational
challenge for direct measurements. Since the background radiation contains 
important information about star formation in the early universe and the 
subsequent evolution of galaxies, the derived constraints on the diffuse
infrared background can have serious cosmological implications.

An early surprise coming out of recent TeV gamma-ray observations is the 
realization that the universe seems to be much more transparent at TeV 
gamma-ray energies than what was previously thought or, equivalently, the 
infrared background is much less intense (Aharonian et al.~\cite{E1101}). 
The results were initially based on gamma-ray observations of two blazars
at moderate redshifts. At the wavelengths of $\sim$1--3 $\mu m$, the derived 
upper limit is barely above the level of integrated light from the galaxies 
that has already been resolved by {\it Spitzer}. This has since been
confirmed by independent measurements (Aharonian et al.~\cite{E0347,E0229};
Albert et al.~\cite{3c279}). 

The results can be enhanced in two ways. One is to extend the spectral 
coverage both to lower and higher energies (beyond 1 TeV), to extend the 
constraints on the IR background over a broader spectral range; the other 
is to observe a large sample of blazars at a range of redshifts, to separate 
intrinsic and extrinsic effects on the gamma-ray spectrum of blazars. The 
latter is obviously also important for understanding gamma-ray production
and propagation in blazars. Meaningful enhancements will likely require a 
significant improvement in the sensitivity of gamma-ray observations, as
well as in the discovering capability of the next-generation of observatories.

On a different front, despite intense observational efforts, the search 
for dark matter signals has not yielded any meaningful constraints on
theoretical models. The detection of TeV gamma rays from the direction of 
the Galactic center (Tsuchiya et al.~\cite{Tsuchiya04}; Kosack et al.~\cite{Kosack04}; Aharonian et al.~\cite{GC}) generated a great deal
of excitement about the prospect of finally seeing, albeit indirect, 
evidence for the annihilation of dark matter particles (Horns~\cite{Horns05}),
since the region 
had been thought of as the best place to search for gamma-ray emission 
resulting from such signals (Berezinsky et al.~\cite{Berezinsky94}; 
Bergstrom et al.~\cite{Berg98,Berg01}; Cesarini et al.~\cite{Cesarini04}; 
Hooper \& Dingus~\cite{Hooper04}). However, it was
quickly realized that more mundane explanations involving SNRs or PWNe
would be more viable (Wang et al.~\cite{Wang06}). This is a generic 
problem with targets (such as the central region of galaxies, clusters of
galaxies, etc.) in which there exist plausible astronomical TeV gamma-ray 
emitters.

A more promising class of systems for indirect dark matter searches are
dwarf spheroidal galaxies (e.g., Gilmore et al.~\cite{Gilmore08}), which 
have much larger mass-to-light ratios than normal galaxies. They have 
indeed become a focus of recent observational efforts (Wood et al.~\cite{Wood08}; Albert et al.~\cite{Draco}; Aliu et al.~\cite{WillmanI}).
To date, no positive detection has been reported.

\subsection{Impact on Cosmic Ray Physics}
\label{subsect:cosray}

The origin of cosmic rays has remained an unresolved issue. For cosmic rays 
below $\sim 10^{15}$ eV, 
it is almost taken for granted that they are associated with SNRs in the
Galaxy. Strong shocks at the outer edge of SNRs are naturally thought of as
the site for particle acceleration. If it is the case, SNRs ought to be among
the most promising targets for TeV gamma-ray experiments. It is reassuring, 
therefore, that an increasing number of SNRs have been detected at TeV 
energies. Unfortunately, this success has not led to a direct proof of the 
production of cosmic rays in SNRs, because, as discussed in 
\S~\ref{subsubsect:snr}, 
the observed gamma rays could be accommodated by either leptonic or
hadronic scenarios. It is, however, hopeful that as the quality of data 
improves, the observations may begin to unveil the characteristics of 
$\pi^0$ decay.

On the other hand, the detection of diffuse TeV gamma-ray emission in the 
Cygnus region and around the Galactic Ridge has provided direct evidence for 
interactions between cosmic rays and molecular clouds in the 
Galaxy (Abdo et al.~\cite{Cygnus}; Aharonian et al.~\cite{GC_ridge}). The 
excellent spatial correlation between the 
Galactic Ridge emission and molecular clouds has left little room
for an alternative explanation. The measured gamma-ray spectrum indicates
that the spectrum of cosmic rays near the center of the Galaxy is 
significantly harder than that in the solar neighborhood, presumably a 
propagation effect (Aharonian et al.~\cite{GC_ridge}. Moreover, the 
density of cosmic
rays seems to be many times the local density. It is argued that the
observations can be explained by the presence of a particle accelerator
near the Galactic center that has been active over the past $10^4$ years. An 
obvious candidate is the supernova remnant Sgr A East, which has about the 
right age. Moreover, Sgr A East is a plausible counterpart of HESS J1745-290, 
whose TeV spectrum has a similar shape to the spectrum of the diffuse emission.

\section{Concluding Remarks}
\label{sect:conclud}

The field of TeV gamma-ray astronomy has matured immensely in recent years 
and has become a viable branch of modern astronomy. It provides a unique 
window to the extreme non-thermal side of the universe. Many classes 
of astronomical systems have been detected at TeV energies. The observations 
have not only shed new light on the properties of the systems themselves but also
on the physical processes operating in diverse astronomical settings. For 
instance, taken together, Blazars, microquasars, and gamma-ray bursts (though
none have been detected at TeV energies yet; Atkins et al.~\cite{Atkins05}; Albert et al.~\cite{MAGICGRB}; Horan et al.~\cite{Horan07}) 
may offer an excellent opportunity
for us to make some tangible comparisons of the processes of particle 
acceleration and interaction in the jets of black holes over a vast range of 
physical scales (from microparsecs to megaparsecs; Cui~\cite{Cui05}). As the 
capability of TeV observatories improves, it is hopeful that more sources in 
the established classes and, more importantly, new classes of sources (e.g., 
GRBs, clusters of galaxies, etc.) will be detected.

The field is of equally great interest to physicists, because it has made it 
possible to study some of the most important questions in physics at energies 
much beyond the capabilities of present and future particle accelerators.
Independent of theoretical scenarios, TeV observations are capable of
constraining the intrinsic spectrum of emitting particles and thus casting
light on the nature of the particles and on the acceleration mechanisms. 
TeV observations have already begun to have a serious impact on modern 
cosmology. They have also provided insights into such fundamental issues as 
dark matter, evaporation of primordial black holes, and test of Lorentz 
invariance. The constraints are expected to improve as the quality of data 
improves. In some cases, however, the challenge is to separate astronomical 
and physical origins of the TeV photons detected.

\section{Future Perspective}

It is perhaps instructive to compare the development of TeV gamma-ray 
astronomy with that of X-ray astronomy. Table~\ref{Tab:milestones} shows
roughly similar stages in the development of the respective fields. We think
that TeV gamma-ray astronomy is roughly where X-ray astronomy was in the 
{\it Einstein} days, when direct fine imaging became possible for the very 
first time. Coincidentally, it took both fields about the same amount of 
time to reach this stage of development, following the detection of the first 
extra-solar source. A noteworthy difference is that an all-sky survey had 
been conducted with the {\it Uhuru} satellite, before the {\it Einstein} 
satellite was put in orbit, at a flux limit that is several orders of 
magnitude below the fluxes of the X-ray sources detected at the time. 
Although a full-sky survey had also been carried out at TeV energies with 
{\em Milagro} (and also {\em Tibet AS$\gamma$}), before HESS came online, the 
flux limit reached is only comparable to the brightest TeV sources known at 
the time (Atkins et al.~\cite{Atkins04}).

The {\em Uhuru} survey provided much needed guidance to subsequent X-ray 
missions. It was superseded in the 90s by a much more sensitive survey 
carried out with the {\em ROSAT} satellite. The {\em ROSAT} survey saw 
nearly all classes of astronomical sources, from stars to AGNs to clusters 
of galaxies, and is truly a key milestone in the development of X-ray 
astronomy. We believe that a similar comprehensive survey is imperative
to the development of TeV gamma-ray astronomy. A glimpse of the importance
of such a survey is provided by the HESS survey of the Galactic central
region (Aharonian et al.~\cite{HGC_1,HGC_2}). Though very limited in
scope, the HESS survey has led to many of the most exciting recent 
discoveries in the field. This implies that a full-sky survey with roughly 
the sensitivity of {\em HESS} would be needed to unveil what is below
the tip of the iceberg already seen. 

Two wide-field surveying experiments have been proposed, based on the water Cherenkov technique that was pioneered by the {\em Milagro} collaboration and was proven to be remarkably successful. The High Altitude Water Cherenkov (HAWC)\footnote{See http://hawc.umd.edu/} experiment is to be located in Sierra Negra, Mexico, which is about 4100 m above the sea level.
In the proposed configuration, {\em HAWC} is expected to be 10-15 times
more sensitive than {\em Milagro}. Being a ground-based experiment, it
would be easily expandable (by adding more water tanks) and thus become
more sensitive. The Large High Altitude Air Shower Observatory (LHAASO)
is still being defined. It is envisioned to spread over an area of one 
square kilometer and to be located at the Yanbajing (YBJ) cosmic ray observatory
in Tibet, China, which is about 4300 m above the sea level. It is worth
noting that YBJ now hosts two on-going experiments, {\em Tibet AS$\gamma$}
and {\em ARGO}. Therefore, the excellent infrastructure is already in place 
for {\em LHAASO}. Since the project is still evolving, the ultimate sensitivity
of {\em LHAASO} is not known at the present time. 
There are also plans to incorporate Cherenkov telescopes into the 
observatory, making it a wide-field and narrow-field combo, very much like,
e.g., {\em RXTE} and {\em Swift}, for X-ray astronomy.
\begin{table}
\begin{center}
\caption[]{Development of X-ray Astronomy and TeV Gamma ray Astronomy.}\label{Tab:milestones}

 \begin{tabular}{lcc}
  \hline\noalign{\smallskip}
Stages & X-ray Astronomy     &  TeV Gamma-ray Astronomy    \\
  \hline\noalign{\smallskip}
initial activities  & sounding rocket and balloon experiments & Cherenkov and air shower experiments  \\ % new variable
first detection  & Sco X-1    &   Crab Nebula              \\
follow-ups  & more detections     &   more detections      \\
first survey & Uhuru  satellite & Milagro	   \\
direct fine imaging & Einstein satellite & HESS, VERITAS	   \\
further development & Ariel 5, ROSAT, ASCA, etc. &  HAWC? LHAASO? CTA? AGIS?  \\
state-of-the-art survey & ROSAT all-sky survey & HAWC? LHAASO? \\
wide- and narrow-field combo & RXTE, Swift & LHAASO? \\
state-of-the-art imaging & Chandra, XMM-Newton & CTA? AGIS? \\ 
  \noalign{\smallskip}\hline
\end{tabular}
\end{center}
\end{table}

Lessons from the success of X-ray astronomy show the importance of parallel 
development of narrow-field imaging and wide-field surveying experiments. The 
two are complementary both in technique (see \S~\ref{subsect:principles}) and in 
scientific capabilities. Narrow-field imaging experiments enable detailed studies of 
TeV gamma-ray sources, while wide-field surveying experiments focus mainly on 
discovering new sources and thus point the way for deep observations with 
narrow-field experiments. In particular, the latter are most effective in catching 
transient phenomena that could be associated with supernova or hypernova 
explosions, merging of neutron stars, nova outburst, blazar outbursts, evaporation 
of primordial black holes, or processes that have not even been thought of yet. 
Rapid extragalactic transient gamma-ray signals have been used as probes 
into some of the most fundamental questions in physics, such as violation of 
Lorentz invariance.  Wide-field surveying experiments can also more easily 
facilitate multi-wavelength observations, which have proven to 
be critical to understanding the processes of particle acceleration and 
radiation production in astronomical environments. 

Compared to space-based experiments (such as {\em Fermi}), ground-based experiments are easily serviceable and can thus potentially run for a long time. This is important for studies that require high statistical precision (as well as a large sample of sources), such as dark matter search. Ground-based experiments can also be upgraded to improve sensitivity. As long as systematic uncertainties can be controlled at a sufficiently low level, a wide-field experiment like {\em LHAASO} has the potential of being an effective pathfinder for the next-generation narrow-field imaging experiments (such as CTA or AGIS). This also represents an excellent opportunity for China to become a major player in the young but exciting field of TeV gamma ray astronomy. To fully expore the promise of wide-field surveying experiments, it would be ideal to have two observatories in the northern and southern hemispheres, respectively, to cover the whole sky. This is certainly another area where international cooperation and collaboration would be critical.

The history of astronomy is full of examples that illustrate how new observational
capabilities bring about new discoveries.  Even in a branch as mature as radio 
astronomy, new transient phenomena are still being discovered. For instance, the 
recent discovery of rotating radio transients (RRATs) shows the presence of an 
intriguing population of radio pulsars that reveal themselves only in sporadic, 
ultra-short radio pulses (which last for milliseconds). In optical astronomy, the 
Sloan Digital Sky Survey (SDSS) has, in many way, changed the way that
research is done. There is no pause in the effort. Many new survey experiments 
are being implemented or proposed, including the Panoramic Survey Telescope 
and Rapid Response System (Pan-STARRS), the Dark Energy Survey (DES), 
and the Large Synoptic Survey Telescope (LSST),  and they promise to
revolutionize the field. At higher energies, the {\em EGRET} survey marked the 
beginning of GeV gamma ray astronomy, which is being further advanced by 
{\em Fermi}. We should not expect TeV gamma ray astronomy to be any 
different --- a strong effort in developing sensitive survey experiments is required
to push the field to the next level.

\section{Acknowledgment}

We wish to thank Felix Aharonian and Peter Biermann for providing valuable comments on the manuscript, and Konrad Bernl\"{o}hr for providing the figures used in Figure~1.  We also acknowledge useful discussions with many participants of the 2008 TeV Particle Astrophysics Workshop in Beijing. This work was partially supported by the US Department of Energy.

\label{lastpage}


\begin{thebibliography}{99}
\bibitem[2007]{Cygnus}Abdo A.A. et al., 2007, \apj, 658, L33
\bibitem[2009a]{Fermi_vela}Abdo A.A. et al., 2009a, \apj, 696, 1084
\bibitem[2009b]{Fermi_cat}Abdo, A.A. et al., 2009b, \apj, in press (arXiv:0902.1340)
\bibitem[2008a]{wcomae}Acciari V. A. et al., (VERITAS Collaboration) 2008a, \apj, 684, L73
\bibitem[2008b]{M87}Acciari V. A. et al., (VERITAS Collaboration) 2008b, \apj, 679, 397
\bibitem[2008c]{LSIver}Acciari V. A. et al., (VERITAS Collaboration) 2008c, \apj, 679, 1427
\bibitem[2009a]{3C66A}Acciari V. A. et al., (VERITAS Collaboration) 2009a, \apj, 693, L104
\bibitem[2009b]{Gall09}Acciari V. A. et al., (VERITAS Collaboration) 2009b, \apj, submitted
\bibitem[2000]{Ahaproton00}Aharonian F., 2000, New Astronomy, 5, 377
\bibitem[2002]{J2032}Aharonian F. et al., 2002, \aap, 393, L37 
\bibitem[2003]{M87hegra}Aharonian F. et al., 2003, \aap, 403, L1
\bibitem[2004a]{J1713}Aharonian F. et al., (HESS Collaboration) 2004a, \nat, 432, 75
\bibitem[2004b]{GC}Aharonian F. et al., (HESS Collaboration) 2004b, \aap, 425, L13 
\bibitem[2005a]{G0.9}Aharonian F. et al., (HESS Collaboration) 2005a, \aap, 432, L25
\bibitem[2005b]{MSH}Aharonian F. et al., (HESS Collaboration) 2005b, \aap, 435, L17
\bibitem[2005c]{HGC_1}Aharonian F. et al., (HESS Collaboration) 2005c, Science, 307, 1938
\bibitem[2005d]{G18}Aharonian F. et al., (HESS Collaboration) 2005d, \aap, 442, L25
\bibitem[2005e]{B1259}Aharonian F. et al., (HESS Collaboration) 2005e, \aap, 442, 1
\bibitem[2005f]{PKS2155}Aharonian F. et al., (HESS Collaboration) 2005f, \aap, 442, 895
\bibitem[2005g]{LS}Aharonian F. et al., (HESS Collaboration) 2005g, Science, 309, 746
\bibitem[2005h]{HESS2032}Aharonian F., et al., (HESS Collaboration) 2005h, A\&A, 431, 197
\bibitem[2006a]{VelaX}Aharonian F. et al., (HESS Collaboration) 2006a, \aap, 448, L43
\bibitem[2006b]{J1713a}Aharonian F. et al., (HESS Collaboration) 2006b, \aap, 449, 223
\bibitem[2006c]{HGC_2}Aharonian F. et al., (HESS Collaboration) 2006c, \apj, 636, 777
\bibitem[2006d]{GC_ridge}Aharonian F. et al., (HESS Collaboration) 2006d, \nat, 439, 695
\bibitem[2006e]{E1101}Aharonian F. et al., (HESS Collaboration) 2006e, \nat, 440, 1018
\bibitem[2006f]{G313}Aharonian F. et al., (HESS Collaboration) 2006f, \aap, 456, 245
\bibitem[2006g]{LSorb}Aharonian F. et al., (HESS Collaboration) 2006g, \aap, 460, 743
\bibitem[2006h]{M87var}Aharonian F. et al., (HESS Collaboration) 2006h, Science, 314, 1424
\bibitem[2006i]{J1825}Aharonian F. et al., (HESS Collaboration) 2006i, \aap, 460, 365
\bibitem[2007a]{J1713b}Aharonian F. et al., (HESS Collaboration) 2007a, \aap, 449, 223
\bibitem[2007b]{PKSflares}Aharonian F. et al., (HESS Collaboration) 2007b, \apj, 664, L71
\bibitem[2007c]{E0347}Aharonian F. et al., (HESS Collaboration) 2007c, \aap, 473, L25
\bibitem[2007d]{E0229}Aharonian F. et al., (HESS Collaboration) 2007d, \aap, 475, L9
\bibitem[2008]{Aharev08}Aharonian F., Buckley J., Kifune T., Sinnis G., 2008, Rep. Prog. Phys., 71, 096901
\bibitem[2009]{cena}Aharonian F. et al., (HESS Collaboration) 2009, \apjl, 965, L40
\bibitem[2006a]{LSI}Albert J. et al., (MAGIC Collaboration) 2006a, Science, 312, 1771
\bibitem[2006b]{MAGICGRB}Albert, J. et al., (MAGIC Collaboration) 2006b, \apj, 641, L9
\bibitem[2007]{m5flares}Albert J. et al., (MAGIC Collaboration) 2007, \apj, 669, 862
\bibitem[2008a]{M87magic}Albert J. et al., (MAGIC Collaboration) 2008a, \apj, 685, L23
\bibitem[2008b]{MAGIC2032}Albert J. et al., (MAGIC Collaboration) 2008b, \apj, 675, L25
\bibitem[2008c]{Draco}Albert J. et al., (MAGIC Collaboration) 2008c, \apj, 679, 428
\bibitem[2008]{crabpulsar}Aliu E. et al., (MAGIC Collaboration) 2008, Science, 322, 1221
\bibitem[2009]{LSIorb}Albert J. et al., (MAGIC Collaboration) 2009, \apj, 693, 303
\bibitem[2009]{WillmanI}Aliu E. et al., (MAGIC Collaboration) 2009, \apj, 697, 1299
\bibitem[2006]{Amenomori06}Amenomori M. et al., 2006, Science, 314, 439
\bibitem[2004]{Atkins04}Atkins R. et al., 2004, \apj, 608, 680
\bibitem[2005]{Atkins05}Atkins R. et al., 2005, \apj, 630, 996
\bibitem[2001]{Bai01}Bai J. M., Lee M. G., 2001, \apj, 549, L173
\bibitem[1999]{Beall99}Beall J. H., Bednarek, W., 1999, \apj, 510, 188
\bibitem[2008]{Begelman08}Begelman M. C., Fabian A. C., Rees M. J., 2008, \mnras, 384, L19
\bibitem[1994]{Berezinsky94}Berezinsky V., Bottino A., Mignola G, 1994, Phys. Lett., B325, 136
\bibitem[1998]{Berg98}Bergstrom L., Ullio P., Buckley, J. H., 1998, Astropart. Phys., 9, 137
\bibitem[2001]{Berg01}Bergstrom L., Edsjo J., Ullio P., 2001, Phys. Rev. Lett., 87, 251301
\bibitem[2005]{Blazejow05}B{\l}a\.zejowski M. et al., (VERITAS Collaboration) 2005, \apj, 630, 130
\bibitem[1996]{BM96}Bloom S. D., Marscher A. P., 1996, \apj, 461, 657
\bibitem[1997]{Boettcher97}B\"ottcher M., et al., 1997, \aap, 324, 395 
\bibitem[2005]{Boettcher05}B\"{o}ttcher M., 2005, \apj, 621, 176
\bibitem[2005]{G12.82}Brogan C. L. et al., 2005, \apj, 629, L105
\bibitem[2005a]{Casares05a}Casares,~J. et al., 2005a, \mnras, 360, 1091
\bibitem[2005b]{Casares05b}Casares,~J. et al., 2005b, \mnras, 364, 899
\bibitem[2004]{Cesarini04}Cesarini A. et al., 2004, Astropart. Phys., 21, 267
\bibitem[2008]{CKC08}Chang C., Konopelko A., Cui W., 2008, ApJ, 682, 1177
\bibitem[2004]{Cui04}Cui W., 2004, \apj, 605, 662
\bibitem[2005]{Cui05}Cui W., 2005, Science, 309, 714
\bibitem[2006]{Cui06}Cui W., 2006, In: Proceedings of The Vulcano Workshop on 
"Frontier Objects in Astrophysics and Particle Physics", eds. F. Giovannelli \& G. Mannocchi, Italian Physical Society, Editrice Compositori, Vol. 93, p. 207 (astro-ph/0608042)
\bibitem[2007]{CK07}Cui W., Konopelko A., 2007, ApJ, 665, L83
\bibitem[1997]{Dar97}Dar A., Laor, A., 1997, \apj, 478, L5
\bibitem[1992]{Dermer92}Dermer C. D. et al., 1992, \aap, 256, L27
\bibitem[2008]{3c279}Errando M. et al., (MAGIC Collaboration) 2008, In: Proceedings of the 4th Heidelberg International Symposium on High Energy Gamma-Ray Astronomy, AIP Conf. Proc., Vol 1085, p. 423 (arXiv:0901.3275)
\bibitem[1996]{Gaidos96}Gaidos J. A. et al., 1996, \nat, 383, 319.
\bibitem[1964]{Ginsburg64}Ginsburg V. L., Syrovatskii, S. I., 1964, In: The Origin of Cosmic Rays (New York: Pergamon Press)
\bibitem[2005]{Ghisellini05}Ghisellini G., Tavecchio F., Chiaberg M., 2005, \aap, 432, 401
\bibitem[2008]{Gilmore08}Gilmore G. et al., 2008, \apj, 663, 948
\bibitem[2002]{gregory02}Gregory P.C., 2002, \apj, 575, 427 
\bibitem[2004]{Hooper04}Hooper D., Dingus B. L., 2004, Phys. Rev. D, 70, 113007
\bibitem[2005]{Horns05}Horns D., 2005, Phys. Lett., B607, 225 (Erratum: B611, 297)
\bibitem[2007]{Horns07}Horns D. et al., 2007, \aap, 469, L17
\bibitem[2007]{Horan07}Horan D. et al., (VERITAS Collaboration) 2007, \apj, 655, 396
\bibitem[2009]{Horan09}Horan D. et al., (VERITAS Collaboration) 2009, \apj, 695, 596
\bibitem[2004]{Kawachi04}Kawachi A. et al., 2004, \apj, 607, 949
\bibitem[2008]{khangulyan08}Khangulyan D., Aharonian F., Bosch-Ramon V., 2008, \mnras, 383, 467
\bibitem[2000]{knodlseder00}Kn\"{o}dlseder J., 2000, \aap, 360, 539
\bibitem[2003]{Konopelko03}Konopelko A. et al., 2003, \apj, 597, 851
\bibitem[2007]{VER2032}Konopelko,~A., et al., (VERITAS Collaboration) 2007, Apj, 658, 1062
\bibitem[2004]{Kosack04}Kosack K. et al., (VERITAS Collaboration) 2004, \apj, 608, L97
\bibitem[2004]{Kraw04}Krawczynski H. et al., 2004, \apj, 601, 151.
\bibitem[2002]{Krennrich02}Krennrich F. et al., 2002, \apj, 575, L9
\bibitem[2008]{Lenain08}Lenain J.-P., Boisson C., Sol H., Katarzy\'{n}ski K., 2008, \aap, 478, 111
\bibitem[2007]{Levinson07}Levinson A., 2007, \apj, 671, L29
\bibitem[2003]{Lyutikov03}Lyutikov M., 2003, New Astr. Rev. 47, 513
\bibitem[1992]{Manheim92}Mannheim~K., Biermann~P.~L., 1992, \aap, 253, L21
\bibitem[1992]{Maraschi92}Maraschi L., Ghisellini G., Celotti A., 1992, \apj, 397, L5
\bibitem[1998]{Marti98}Mart\'{\i} J., Paredes J. M., Rib\'o, M., 1998, A\&A, 338, L71
\bibitem[1997]{MK97}Mastichiadis A., Kirk J. G., 1997, \aap, 320, 19
\bibitem[1997]{Motch97}Motch C. et al. 1997, A\&A, 323, 853
\bibitem[2003]{mucke03}M\"ucke A. et al., 2003, Astropart. Phys., 18, 593
\bibitem[2000]{Paredes00}Paredes J. M., Mart\'{\i} J., Rib\'o M., Massi M., 2000, Science, 288, 2340
\bibitem[2000]{Pohl00}Pohl M., Schlickeiser R., 2000, \aap, 354, 39
\bibitem[2003]{Protheroe02}Protheroe R. J., Donea A.-C., Reimer A., 2003, Astropart. Phys., 19, 559
\bibitem[1978]{Rees78}Rees M. J., 1978, \mnras, 184, 61 
\bibitem[1999]{Ribo99}Rib\'o M., Reig P., Mart\'{\i} J., Paredes J. M., 1999, A\&A, 347, 518
\bibitem[2009a]{SierTorres09a}Sierpowska-Bartosik A., Torres D.F., 2009a, \apj, 693, 1462
\bibitem[2009b]{SierTorres09b}Sierpowska-Bartosik A., Torres D.F., 2009b,  Astroparticle Physics, in press (arXiv:0801.3427)
\bibitem[1994]{Sikora94}Sikora M. et al., 1994, \apj, 421, 153
\bibitem[2001]{Spada01}Spada M., et al., 2001, \mnras, 325, 1559 
\bibitem[2006]{Stawarz06}Stawarz L. et al., 2006, \mnras, 370, 981
\bibitem[2008]{Tanaka08}Tanaka T. et al., 2008, \apj, 685, 988
\bibitem[2008]{Tavecchio08}Tavecchio F., Ghisellini G., 2008, \mnras, 385, L98
\bibitem[2008]{J1731}Tian W., Leahy D. A., Haverkorn M., Jiang B., 2008, \apj, 679, L85
\bibitem[2004]{TDR04}Torres D.F., Domingo-Santamaria, E., Romero, G.E., 2004, \apj, 601, L75
\bibitem[2004]{Tsuchiya04}Tsuchiya K. et al., 2004, \apj, 606, L115
\bibitem[2007]{Uchiyama07}Uchiyama Y. et al., 2007, Nature, 449, 576
\bibitem[1995]{urry95}Urry C. M., Padovani P., 1995, \pasp, 107, 803
\bibitem[2006]{Wang06}Wang Q. D., Lu F. J., Gotthelf E. V., 2006, \mnras, 367, 937
\bibitem[1989]{Weekes89}Weekes T. C. et al., 1989, \apj, 342, 379
\bibitem[1992]{WC92}White R., Chen W., 1992, ApJ, 387, L81
\bibitem[2008]{Wood08}Wood M., et al., (VERITAS Collaboration)  2008, \apj, 678, 594
\bibitem[2005]{Xue05}Xue Y. Q., Cui W., 2005, \apj, 622, 160
\bibitem[2006]{XYC06}Xue Y. Q., Yuan, F., Cui W., 2006, \apj, 647, 194
\end{thebibliography}
\end{document}